\documentclass[11pt]{article}
\pdfoutput=1
\usepackage{amsmath}
\usepackage{amssymb}
\usepackage{graphicx,bbm,mathrsfs}
\usepackage{nicefrac}
\usepackage{slashed}
\usepackage{bbm}
\usepackage{geometry}
\usepackage{empheq}
\usepackage{stackrel}
\usepackage{ulem}
\usepackage{xcolor}

\usepackage{jheppub}
\usepackage{setspace}
\usepackage{relsize}
\usepackage{empheq}
\usepackage{wasysym}
\usepackage{hhline,colortbl}
\usepackage{pifont}
\newcommand{\cmark}{\ding{51}}%
\newcommand{\xmark}{\ding{55}}%

\usepackage{mathtools}
\usepackage{enumitem}
\usepackage{dcolumn}   
\usepackage{bm}        
\usepackage{graphicx,mathrsfs}
\usepackage{nicefrac}
\usepackage{multirow}
\usepackage{color}
\usepackage{mathtools}
\usepackage{makecell}
\usepackage{environ} 
\usepackage{lipsum}

 \NewEnviron{Smaller11}{
           \scalebox{1.1}{$\BODY$} 
 } 
 \NewEnviron{Smaller08}{
           \scalebox{0.8}{$\BODY$} 
 }

\newcommand{\be}{\begin{equation}}
\newcommand{\ee}{\end{equation}}
\newcommand{\bea}{\begin{eqnarray}}
\newcommand{\eea}{\end{eqnarray}}

\renewcommand{\S}{\cal S}
\newcommand{\M}{\cal M}
\newcommand{\Mnu}{{\cal M}_\nu}
\newcommand{\MnuM}{{\cal M}_\nu^-}
\newcommand{\MnuP}{{\cal M}_\nu^+}

\newcommand{\x}{\mathbf{x}}

\newcommand{\Y}{\mathcal Y}

\renewcommand{\S}{{\cal S}}

\newcommand{\h}{h}

\usepackage{titlesec}

\titleformat*{\section}{\Large\bfseries}
\titleformat*{\subsection}{\large\bfseries}
\titleformat*{\subsubsection}{\large\bfseries}
\titleformat*{\paragraph}{\large\bfseries}
\titleformat*{\subparagraph}{\large\bfseries}

\makeatletter
\newcommand*{\prodsym}{%
  \DOTSB
  \mathop{
    \mathchoice
      {\rlap{\kern.3em\rotatebox[origin=c]{-90}{}}{\prod}}
      {\vcenter{\rlap{\kern.2em\rotatebox[origin=c]{-90}{}}}{\prod}}
      {\sum}{\sum}
  }\slimits@
}
\makeatother

\makeatletter
\DeclareFontFamily{OMX}{MnSymbolE}{}
\DeclareSymbolFont{MnLargeSymbols}{OMX}{MnSymbolE}{m}{n}
\SetSymbolFont{MnLargeSymbols}{bold}{OMX}{MnSymbolE}{b}{n}
\DeclareFontShape{OMX}{MnSymbolE}{m}{n}{
    <-6>  MnSymbolE5
   <6-7>  MnSymbolE6
   <7-8>  MnSymbolE7
   <8-9>  MnSymbolE8
   <9-10> MnSymbolE9
  <10-12> MnSymbolE10
  <12->   MnSymbolE12
}{}
\DeclareFontShape{OMX}{MnSymbolE}{b}{n}{
    <-6>  MnSymbolE-Bold5
   <6-7>  MnSymbolE-Bold6
   <7-8>  MnSymbolE-Bold7
   <8-9>  MnSymbolE-Bold8
   <9-10> MnSymbolE-Bold9
  <10-12> MnSymbolE-Bold10
  <12->   MnSymbolE-Bold12
}{}

\let\llangle\@undefined
\let\rrangle\@undefined
\DeclareMathDelimiter{\llangle}{\mathopen}%
                     {MnLargeSymbols}{'164}{MnLargeSymbols}{'164}
\DeclareMathDelimiter{\rrangle}{\mathclose}%
                     {MnLargeSymbols}{'171}{MnLargeSymbols}{'171}
\makeatother

\begin{document}

\vspace*{4mm}

\thispagestyle{empty}

\begin{center}

\begin{minipage}{20cm}
\begin{center}
\hspace{-5cm }
\huge
\sc
Holographic Fluids \\
\hspace{-5cm }
from 5D Dilaton Gravity
\end{center}
\end{minipage}
\\[30mm]

\renewcommand{\thefootnote}{\fnsymbol{footnote}}

{\large  
Sylvain~Fichet$^{\, a}$ \footnote{sylvain.fichet@gmail.com}\,, 
Eugenio~Meg\'{\i}as$^{\, b}$ \footnote{emegias@ugr.es}\,,
Mariano~Quir\'os$^{\, c}$ \footnote{quiros@ifae.es}\,
}\\[12mm]
\end{center} 
\noindent

${}^a\!$ 
\textit{Centro de Ciencias Naturais e Humanas, Universidade Federal do ABC,} \\
\indent \; \textit{Santo Andre, 09210-580 SP, Brazil}

${}^b\!$ 
\textit{Departamento de F\'{\i}sica At\'omica, Molecular y Nuclear and} \\
\indent \; \textit{Instituto Carlos I de F\'{\i}sica Te\'orica y Computacional,} \\
\indent \; \textit{Universidad de Granada, Avenida de Fuente Nueva s/n, 18071 Granada, Spain}

${}^c\!$  
\textit{Institut de F\'{\i}sica d'Altes Energies (IFAE) and} \\
\indent \; \textit{The Barcelona Institute of  Science and Technology (BIST),} \\
\indent \; \textit{Campus UAB, 08193 Bellaterra, Barcelona, Spain}

\addtocounter{footnote}{-1}

\vspace*{10mm}
 
\begin{center}
{  \bf  Abstract }
\end{center}
\begin{minipage}{15cm}
\setstretch{0.95}

We study a solvable class of five-dimensional dilaton gravity models that continuously interpolate between anti-de Sitter  (AdS$_5$), linear dilaton (LD$_5$) and positively curved spacetimes 
as a function of a continuous  parameter $\nu$. 
The dilaton vacuum expectation value is set by a potential localized on a flat brane. 
We chart the elementary properties  of these backgrounds for any admissible $\nu$,  and determine  stability conditions of the brane-dilaton system. We find that the spectrum of metric fluctuations  can be either continuous or discrete. It features a massless graviton mode confined between the brane and the curvature singularity, and a massive radion mode  tied to brane-dilaton stability.  We  show that, in the presence of a bulk black hole, the holographic theory living on the brane features a perfect fluid. The equation of state of the holographic fluid   interpolates between radiation, pressureless matter and vacuum energy as a function of $\nu$. 
This extends earlier findings on holographic fluids.
Our results suggest that the thermodynamics  of the  fluid mirrors precisely the thermodynamics of  the bulk black hole.

    \vspace{0.5cm}
\end{minipage}

\newpage
\setcounter{tocdepth}{2}
\tableofcontents

\section{Introduction \label{se:intro}}

Both anti-de Sitter (AdS) and linear dilaton (LD) spacetimes are holographic. 
Observables defined in a given asymptotic region of these spacetimes can, at least for certain dimensions, be equivalently described by a dual theory with  one dimension less and no gravity. 
This fact is obtained from string theory by taking appropriate decoupling limits of certain brane configurations. 
For AdS, this leads to the string-derived AdS$_5$/CFT$_4$ duality \cite{Aharony:1999ti}, while the seven-dimensional linear dilaton background (LD$_7$) is dual to little string theory (LST)
\cite{Seiberg:1997zk,Berkooz:1997cq}.\,\footnote{
LST is a 6D string theory on a stack of {NS5} branes with $g_s\to 0$ \cite{Aharony:1998ub}.
It is an interacting theory of noncritical strings which is nonlocal,  has no massless graviton, and has Hagedorn density of states at high energy \cite{Aharony:1999ks,Kutasov:2001uf}.   Lower-dimensional versions of this LD$_7$/LST$_6$ duality can be obtained via spatial compactifications, {\textit{e.g.}~a simple toroidal $\mathbb T_2$ 
compactification gives rise to the phenomenological theories $\textrm{LD}_5/\textrm{LST}_4$, see \textit{e.g.} \cite{Antoniadis:2021ilm}. 
}
}

At subPlanckian energies,  holography of the AdS and LD spacetimes  can alternatively be explored using  general relativity and effective field theory (EFT). 
A successful approach is to place a brane in the spacetime  and to  study observables defined on this brane.  The bulk spacetime, possibly featuring a black hole, is integrated out. 

This low-energy approach to holography gives compelling results in AdS background, in which case the brane can be seen as a regularized version of the AdS conformal boundary.
At low-energy, the effective theory  of  gravity on the brane is Einstein gravity coupled to a  perfect  fluid that nontrivially emerges from  the bulk physics. This \textit{holographic fluid} turns out to be conformal, as expected from the AdS/CFT correspondence\,\cite{Gubser:1999vj, Shiromizu:1999wj,Binetruy:1999hy,Hebecker:2001nv,Langlois:2002ke,Langlois:2003zb}.

The same  approach also gives compelling results in the LD background. It was shown in \cite{Fichet:2023xbu}  that the holographic fluid emerging on the brane has Hagedorn thermodynamics for any dimension $d$. The  low-energy framework thus reproduces the thermodynamic behavior of  LST 
for $d=6$. It also reproduces the thermodynamics  of  $T\bar T$-deformed CFT  proposed in \cite{Giveon:2017nie} 
for $d=2$.\,\footnote{
 Further evidence for the proposed duality has been developed in  \cite{Giribet:2017imm, Asrat:2017tzd, Araujo:2018rho,Chakraborty:2020fpt,Chakraborty:2020yka,Georgescu:2022iyx, Chang:2023kkq,Chakraborty:2023mzc,Chakraborty:2023zdd,Aharony:2023dod}. }

In this manuscript we extend the above low-energy approach to holography to a class of dilatonic spacetimes that contains the AdS and LD ones as particular cases. 
One motivation for this analysis is to study the behavior of the holographic fluid that emerges on the brane. Since meaningful results are obtained from both AdS and LD, we may wonder how does the  holographic fluid behaves in our more general spacetime.

 Another motivation comes from the fact that the linear dilaton background is  special because it features some simple conformal symmetries.  
While these are much weaker than the AdS isometries, they are nevertheless expected to have physical consequences \cite{Fichet:2023xbu}.
Placing the LD in a broader context helps understanding why physics on the LD background is special. 

An extra motivation for our model is the remarkable simplicity of its solutions. 
 Both classical solutions of the dilaton-gravity system and the quantum fluctuations of gravity are described by compact analytical expressions.  Even though this technical feature is conceptually irrelevant, it is certainly important for practical purposes:  
our  model provides an avenue to easily explore the physics of dilatonic spacetimes.

We present the model and discuss solutions with and without black hole in section~\ref{se:model}. We study the fluctuations of the bulk metric around these solutions in section \ref{se:fluc}.
We compute the holographic fluid and its thermodynamic properties  in section~\ref{se:fluid}. {There we also briefly  discuss a class of braneworld scenarios.} A summary is given in section~\ref{se:summary}.
{The appendices contain details about gauge fixing in holographic actions (App.\,\ref{app:gauge-fixing}), a full calculation of the graviton quadratic action showing explicitly that the 5D graviton is massless (App.\,\ref{app:quadratic_action}), and a discussion of the equations of motion in the scalar sector (App.\,\ref{app:EoM_canonical_radion}). }

\section{A Solvable Dilaton-Gravity Model }
\label{se:model}

The action of the 5D dilaton gravity system {in the Einstein frame} is
\begin{eqnarray}
\S &=& \int d^5x  \sqrt{g}  \bigg(  \frac{M_5^3}{2} {}^{(5)}R  - \frac{1}{2} (\partial_M \phi)^2 - V(\phi)  \bigg) \nonumber \\ 
&&-\int_{\textrm{brane}} d^4x \sqrt{\bar g} \left(V_b(\phi) + \Lambda_b - M_5^3 K \right) ~ + ~ \S_{\textrm{matter}}  
\,. \label{eq:action}
\end{eqnarray}
 ${}^{(5)}R$ is the  scalar curvature, $\phi$ is the dilaton field, $M_5$ is the fundamental 5D Planck scale. 
The spacetime supports a $3$-brane with induced metric  $\bar g_{\mu\nu}$. The brane supports  a tension $\Lambda_b$ and a localized potential $V_b(\phi)$.  $K$ is the extrinsic curvature that appears in  the Gibbons-Hawking-York (GHY) boundary term. 
The $V_b(\phi)$ potential stabilizes  $\phi$ to the vacuum expectation value (vev) $\langle\phi\rangle_{\rm brane}\equiv  v_b$, which  determines completely the background. $\S_{\textrm{matter}}$ encodes the quantum fields living on this background. 

The $V_b(\phi)$ potential does not need to be explicitly specified.
 We use the convention $V_b(v_b)=0$ wihout loss of generality.
The potential and the stability of the dilaton-brane system are  discussed in sections \ref{se:stab} and \ref{se:fluc}, in which further conditions on $V_b$ are derived.

The 5D metric is set to the {general} ansatz
\begin{equation}
ds^2_{\rm gen} = g_{MN} dx^M dx^N  = e^{-2A(r)}\left( - f(r) d\tau^2 + d \x^2 \right) +\frac{e^{-2B(r)}}{f(r)}dr^2   \label{eq:ds2brane}
\,.
\end{equation}
 5D coordinates  are labeled by uppercase Latin indexes $(M,N,\cdots)$, 
4D  coordinates  along the constant-$r$ slices are labeled by Greek indexes $(\mu,\nu,\cdots)$, \textit{i.e.}~$x^M=(x^\mu,r)=(\tau,\x,r)$. 
In the general ansatz \eqref{eq:ds2brane}, we allow {for} a blackening factor $f(r)$ that describes a black hole horizon at the hypersurface $r=r_h$ if  $f(r_h)=0$, and $A(r)$, $B(r)$ and $\phi(r)$ are regular at $r_h$. 
 We assume a flat brane lying at the location $r=r_b$, 
consistently with the Poincaré invariance of the constant-$r$ slices of \eqref{eq:ds2brane},

We introduce the reduced bulk potential $V(\phi) \equiv  3 M_5^3 \bar V(\bar\phi)$ and the reduced dilaton field $\phi \equiv  \sqrt{3 M_5^3} \bar \phi $. $\bar V$  {has mass dimension 2}, {while $\bar\phi$ is dimensionless}. 
The general model of this work is defined
by setting the reduced bulk potential to
\begin{equation}
\bar V(\bar\phi) = - \frac{1}{2}(4-\nu^2) k^2 e^{2 \nu \bar\phi} \,. \label{eq:barV}
\end{equation}
Here  $\nu$ is a real parameter that we take positive without loss of generality.
Further restrictions on $\nu$ are found in subsections \ref{se:M_noBH} and \ref{se:M_BH}.

The field equations are 
{
\begin{eqnarray}
  0 &=& {}^{(5)}R_{MN} - \frac{1}{2} g_{MN} {}^{(5)}R - 3\, \partial_M \bar\phi  \partial_N \bar\phi 
   + \frac{3}{2} g_{MN} (\partial_A\bar\phi)^2 + 3\, g_{MN} \bar V(\bar\phi)  \,, \label{eq:Eg}\\
  0 &=&  \frac{1}{\sqrt{g}} \partial_M\left(\sqrt{g} g^{MN}\partial_N  \bar\phi \right) - \frac{\partial \bar V}{\partial \bar\phi} \,. \label{eq:Ephi}
  \end{eqnarray}
}
The solutions to the field equations have some integration constants that need a careful analysis. Some are gauge redundancies, other are physically meaningful. We refer to \cite{Fichet:2023xbu} for a detailed discussion.

A combination of integration constants is fixed by the fact that the value of $v_b$ does not change with the brane location,  $\frac{\partial v_b}{\partial r_b}=0$ \cite{Fichet:2023xbu}. This is a  natural consequence of the fact that the $V_b$ potential is independent on the brane location. 
Upon reduction of the integration constants, a physical mass scale $\eta$  appears,  
\be
\eta\equiv k \, e^{\nu\, \bar v_b}\,. \label{eq:eta_def}
\ee

The spacetime manifold that solves the field equations \eqref{eq:Eg} and \eqref{eq:Ephi} is denoted by~$\Mnu$. The brane at $r=r_b$ partitions $\M_\nu$ into two regions
\be
\MnuM=\Mnu \big|_{r\in(0,r_b]} \,,\quad\quad \MnuP=\Mnu \big|_{r\in[r_b,\infty)} \,. 
\label{eq:Mnu_halves}
\ee
All the quantum fields, including gravitons, have boundary conditions on the brane. 
The quantum fields living in   $\MnuM$ and $\MnuP$ are thus independent of each other.

\paragraph{The $\mathbb{Z}_2$ orbifold convention.}

The fundamental domains of $\MnuM$, $\MnuP$ are $(0,r_b]$ and $[r_b,\infty)$. 
To ease comparison with the extradimensional literature, we find it convenient to adopt the orbifold convention for which space is mirrored on each side of the brane. In this paper, the brane-localized quantities $\Lambda_b$, $V_b$ are  defined within this convention. 
If,  instead, one chooses to define the action only on the fundamental domain, $S_{\rm bulk}$ and $S_{\rm GHY}$ both get reduced by a factor of two, and
all the subsequent computations are equivalent to those made in the orbifold convention upon  changing the definition of the brane localized quantities as $\Lambda_b\to \frac{1}{2}\Lambda_b $, $V_b\to \frac{1}{2}V_b$.

\subsection{The $\Mnu$ Spacetime with No Black Hole  }
\label{se:M_noBH}

In the absence of a black hole {(\textit{i.e.}~Eq.~(\ref{eq:ds2brane}) with $f(r)\equiv 1$),} the solutions to \eqref{eq:Eg} and \eqref{eq:Ephi}  are
\be
ds^2_{\nu,r_b} = \left(\frac{ r}{L}\right)^2 \eta_{\mu\nu}dx^\mu dx^\mu + 
\left(\frac{  r}{r_b}\right)^{2\nu^2}
\frac{1}{(\eta r)^2}
d r^2\,, \label{eq:ds2_Mnu}
\ee
\be
\bar\phi(r) = \bar \phi_b - \nu \log\left( \frac{r}{r_b} \right)  \,, \label{eq:barphi_r}
\ee
with $r\in\mathbb{R}_+$. 
In \eqref{eq:barphi_r}, $\bar\phi_b =\frac{1}{\sqrt{3M_5^3}} \phi_b$ is the value of the reduced dilaton field on the brane, with $\phi_{\rm brane}\equiv\phi_b$. The value  of $\phi_b$
 is set to the vev $\langle\phi_b\rangle =v_b$ due to the brane-localized potential.

The nonzero components of the Ricci tensor are
\begin{equation}
 R_{ii} =  - R_{\tau \tau} = (  \nu^2-4) \eta^2 \left( \frac{r_b}{L}\right)^2 \left( \frac{r}{r_b} \right)^{2(1- \nu^2)} \,, \qquad
R_{rr} = \frac{4 (\nu^2 - 1)}{r^2} \,, \label{eq:Ricci}
\end{equation}
and the scalar curvature is 
\be
R = 4 (2\nu^2 - 5) \eta^2 \left( \frac{r_b}{r} \right)^{2 \nu^2} \,. \label{eq:R}
\ee
The scalar curvature is negative, zero and positive for $\nu<\nu_c$, $\nu=\nu_c$, $\nu>\nu_c$, respectively, with $\nu_c \equiv \sqrt{5/2}$. ${\cal M}_{\nu_c}$ is however not Ricci-flat, as can be seen from \eqref{eq:Ricci}. Some noticeable cases appear:
\begin{itemize}

\item   For $\nu=0$, the scalar curvature is a negative constant. Thus ${\cal M}_0$ is  AdS$_5$ spacetime. 

\item For $\nu=1$, the dilatation $r\to\lambda r$ is a conformal symmetry of $ds^2_{1,r_b}$.
Thus ${\cal M}_1$ is  LD$_5$ spacetime~\cite{Fichet:2023xbu}. This symmetry  implies that the $R_{rr}$ component vanishes. 

\item For $\nu=2$,  only $R_{rr}$ is nonzero. The bulk potential defined in \eqref{eq:barV} vanishes identically. Yet, the spacetime is non-flat  due to the nonzero dilaton kinetic term,  on which the metric backreacts.

\end{itemize}

Importantly, the metric \eqref{eq:ds2_Mnu} depends on the brane location $r_b$, except for $\nu=0$, \textit{i.e.}~AdS. This is because the dilaton vev is fixed on the brane. {When} the brane location varies, the whole spacetime varies accordingly.

\subsubsection{Conformal frame}

We  introduce conformal coordinates.  For any $\nu\neq1$ we have
  {
  \be
   z=\frac{L}{\eta\, r_b} \frac{1}{|\nu^2-1|}\left(\frac{r}{r_b}\right)^{\nu^2-1}\,, \qquad \nu\neq 1 \,,
  \ee
  }
with the domain $z\in \mathbb{R}_+$. 
In conformal coordinates the metric reads
  {
  \be
  ds^2_{\nu,r_b} = \left(\frac{r_b}{L}\right)^{\frac{2\nu^2}{\nu^2-1}}  \left(|\nu^2-1|\eta z\right)^{\frac{2}{\nu^2-1}}
  (\eta_{\mu\nu}dx^\mu dx^\mu  +dz^2)\,.
  \ee
  }
In these coordinates, the $\MnuM$ and $\MnuP$ subspaces defined in \eqref{eq:Mnu_halves} are given by 
\begin{align}
\MnuP&=\Mnu \big|_{z\in(0,z_b]} \,,\quad \MnuM=\Mnu \big|_{z\in[z_b,\infty)} \,\quad {\rm if}\quad \nu < 1   \,,
\nonumber \\
\MnuM &=\Mnu \big|_{z\in(0,z_b]} \,,\quad \MnuP = \Mnu \big|_{z\in[z_b,\infty)} \,\quad {\rm if}\quad \nu > 1 \,,
\label{eq:M_cases}
\end{align}
where
\be
z_b=\frac{L}{\eta\, r_b} \frac{1}{|\nu^2-1|}\,.
\ee

The special case $\nu=1$ is the linear dilaton spacetime, for which 
\be z= \pm \frac{L}{r_b\eta}\log\frac{r}{L} \label{eq:zLD} \,.\ee 
Importantly, the domain in this case is $z\in \mathbb{R}$.  The freedom of sign in \eqref{eq:zLD}  is reminiscent of a discrete  symmetry of the LD spacetime pointed out in \cite{Fichet:2023xbu}.  

\subsubsection{Global properties}

\paragraph{Singularity.} The scalar curvature diverges at $r\to0$ for any  $\nu>0$.  
There is thus a curvature singularity at $r=0$, which  lies in the $\MnuM$ part of the spacetime.  The $\MnuP$ spacetime does not feature any singularity for any $\nu$. 
The singularity is labeled as ``good'' in the sense of Refs.~\cite{Gubser:2000nd,Cabrer:2009we} if $\nu<2$.   As a matter of fact, we will see below that $\nu\in[0,2)$ is the range of values for which the singularity can get censored by a black hole horizon.

\paragraph{Boundaries.}

Using the conformal coordinates with $z\in(0,\infty)$, we see that  if $\nu\in[0,1)$,
$\Mnu$ has a conformal boundary.\,\footnote{That is, 
$\Mnu$ is conformally equivalent to a spacetime with boundary. In our case, it is  for example equivalent to half-Minkowski space. } This boundary is in $\MnuP$. 
For $\nu=1$ (the LD$_5$ spacetime), there is no boundary since $z\in \mathbb{R}$, hence  the LD$_5$ space has the same causal structure as Minkowski space \cite{Fichet:2023xbu}.
Finally for $\nu>1$, there is a regular (\textit{i.e.}~not conformal) boundary at $z=0$. This boundary is in $\MnuM$. {It coincides with the curvature singularity.}

\subsection{Holographic Effective Potential and Stability} 

\label{se:stab}

The $\Mnu$ solutions assume that the brane lies at an arbitrary location $r_b$. Here we determine  under which conditions  this assumption is valid.

We put the  classical solutions \eqref{eq:ds2_Mnu}, \eqref{eq:barphi_r} into the action $\S$. This defines a ``holographic''  on-shell action, $\S_{\rm on-shell} $, that depends only on the   brane location $r_b$ and on the value of the dilaton field on the brane, $\phi_b$. Restricting $\phi_b$ to $x^\mu$-independent configurations, we obtain 
the effective potential
\be 
\S_{\rm on-shell}(r_b,\phi_b) \equiv - \int d^4x \, V_{\rm eff}(r_b,\phi_b) \,
\ee 
that we can use to study stability of the brane-dilaton system.

The on-shell action receives contributions from the bulk, brane and GHY actions, $\S = \S_{\rm bulk} + \S_{\rm brane} + \S_{\rm GHY}$.
The $r$-integral in $\S_{\rm bulk}$  has to be performed. 
The integral in the $\MnuM$ region is finite. The integral in $\MnuP$ is infinite, in that case one regularizes it with a 4D cutoff surface $\Sigma$ located at $r_\Sigma\gg r_b$. Placing a counterterm on $\Sigma$ produces a finite, renormalized on-shell action 
(see \textit{e.g.}~\cite{Mann:2009id}). The counterterm in our case is 
$\S_{\textrm{ct}} = 2M_5^3 \int_{\Sigma} d^4 x \sqrt{h} \, \eta(\phi_\Sigma)$ with $h_{\mu\nu}$ the induced metric on $\Sigma$. 

We find 
\begin{eqnarray}
\S_{\rm bulk} &=&   \mp 2 \int d^4x \, M_5^3 \eta(\phi_b) \left( \frac{r_b}{L} \right)^4 \,,  \label{eq:S_bulk_1} \\
\S_{\rm brane} &=&  - \int d^4x \,(V_b(\phi_b) + \Lambda_b) \left( \frac{r_b}{L} \right)^4 \,, \label{eq:S_brane_1} \\
\S_{\rm GHY} &=&  \pm 8 \int d^4x \, M_5^3 \eta(\phi_b) \left( \frac{r_b}{L} \right)^4 \,, \label{eq:S_GHY_1}
\end{eqnarray}
{for the $\mathcal M^{\mp}_\nu$ spacetimes.} We have introduced the $\phi_b$-dependent mass scale $\eta(\phi_b) = k \, e^{\nu \bar \phi_b}$.
Combining the terms,  the effective potential is 
\begin{equation}
V_{\rm eff}(r_b,\phi_b) = U_b(\phi_b) \left( \frac{r_b}{L} \right)^4 \,,
\end{equation}
where we have defined the \textit{brane-localized effective potential}\,\footnote{ We may recognize the $-6 M_5^3 \eta(\phi_b)$ term as the superpotential of the model, $W(\phi_b) = -6 M_5^3 \eta(\phi_b)$. The bulk potential itself can in general be  expressed as a function of the superpotential as { $V(\phi)= \frac{1}{8} \left( W^{\prime}(\phi)^2 - \frac{4}{3 M_5^3} W^2(\phi) \right) \,$\cite{Fichet:2023xbu}.}  In our case, this gives rise to the bulk potential of Eq.\,\eqref{eq:barV}.  }
\begin{equation}
U_b(\phi_b)=V_b(\phi_b)+\Lambda_b\mp 6 M_5^3 \eta(\phi_b) \quad \textrm{for}\quad \mathcal M_\nu^\mp \,. \label{eq:Ub_def}
\end{equation}

Let us consider the $\phi_b$ direction for any $r_b\neq0$. The potential stabilizes $\phi_b$ to the vev $\langle \phi_b\rangle=v_b$ when the $\phi_b$ derivative vanishes, which corresponds to the condition $U^\prime(v_b) = V_b'(v_b) \mp 2\sqrt{3 M_5^3} \nu \eta(v_b) =0$ for $\mathcal M_\nu^\mp$.  
Stability along the $\phi_b$ direction is ensured if the second $\phi_b$ derivative is positive, \textit{i.e.}
\be
U_b^{\prime\prime}(v_b)>0\,.  \label{eq:U_b_cond}
\ee
Using that 
$\partial_{\phi_b}^2 \eta = \frac{\nu^2}{3M^3_5}  \eta $, 
it follows that for any $r_b\neq0$,  the brane potential $V_b$ must satisfy 
\be
V_b^{\prime\prime}(v_b) > \pm 2 \nu^2 \eta \quad \textrm{for}\quad \mathcal M_\nu^\mp \,, \label{eq:V_b_cond}
\ee
with $\eta(v_b)=\eta$. This is true for any potential $V_b$, we do not need to specify it further than the above stability condition.
Notice that in the {$\MnuM$} space, depending on the form of $V_b$, the $\phi_b=v_b$ minimum may be local if  $|\eta|\gg V_b$ at  large values of $\phi_b$. In such a case,  $V_{\rm eff}$ gets unbounded from below, the $\phi_b=v_b$  vacuum is then metastable.

We  then analyze the $r_b$ direction of $V_{\rm eff}$. Recall we use the convention $V_b(v_b)=0$. When the dilaton is stabilized at  $\phi_b=v_b$,  it turns out that $V_{\rm eff}$ has a minimum at $r_b = 0$ for $ \pm 6 M_5^3 \eta <  \Lambda_b$ for $\mathcal M_\nu^\mp$, and is unbounded from below otherwise. Neither of these cases would be compatible with a brane living at arbitrary values of $r$. 
However, we can assume that $\Lambda_b$ is tuned as 
\be
 \Lambda_b = \pm 6 M_5^3 \eta   \quad \textrm{for}\quad \mathcal M_\nu^\mp \,.\label{eq:Lambda_b_tuned}
\ee
In that case the potential has a flat direction in $r_b$, and the brane can stay at any value of~$r$.  

In summary, we find that the $\Mnu$ solutions are stable provided that the conditions \eqref{eq:V_b_cond} and \eqref{eq:Lambda_b_tuned} are satisfied. 
In section \ref{se:fluc}, the stability condition \eqref{eq:V_b_cond} shows up in a different way by analysis of the metric fluctuations.
In section \ref{se:fluid}, we find that the tuning \eqref{eq:Lambda_b_tuned} corresponds precisely to tuning the 4D cosmological constant to zero in the 4D holographic theory.

\subsection{The $\Mnu$ Spacetime with a Black Hole  }

\label{se:M_BH}

The metric in the presence of a planar black hole is given by 
\be
ds^2_{\nu,r_h,r_b} = \left(\frac{ r}{L}\right)^2 (-f(r)d\tau^2+ d\x^2) + 
\frac{1}{f(r)}\left(\frac{  r}{r_b}\right)^{2\nu^2}
\frac{1}{(\eta r)^2}
d r^2\,, \label{eq:ds2_BH} 
\ee
with
\be
f(r) = 1 - \left( \frac{r_h}{r} \right)^{4-\nu^2} \,. \label{eq:f_r} 
\ee
We assume that the brane is  outside the black hole, \textit{i.e.}~$r_h < r_b$.

We see  from \eqref{eq:ds2_BH}, \eqref{eq:f_r} that the black hole solution exists if $\nu\in[0,2)$. In this range, 
the black hole {interior} is at $r\leq r_h$. Therefore the horizon censors the singularity at $r=0$. This is another way to see that the singularity is good on $\nu\in[0,2)$. The black hole is always in $\MnuM$.

At $\nu=2$,  $f$ vanishes identically, which is inconsistent and means that the black hole cannot exist. Finally, for $\nu>2$, the singularity would be naked. We thus restrict our analysis to 
$\nu\in[0,2)$.

For $\nu\in[0,2)$,  the Hawking temperature  of the black hole horizon is \cite{Fichet:2023xbu}
\begin{equation}
T_h = \frac{4 - \nu^2}{4} \frac{\eta}{\pi} \frac{r_h}{L} \left( \frac{r_b}{r_h} \right)^{\nu^2} \,. \label{eq:Th}
  \end{equation}
We have $T_h\propto r_h$ if $\nu=0$ (\textit{i.e.}~AdS$_5$) and $T_h\propto r_b$ if $\nu=1$ (\textit{i.e.}~LD$_5$).  

The entropy of the black hole can be computed by using the Bekenstein-Hawking entropy formula~\footnote{We have introduced an extra factor of $2$ due to the $\mathbb Z_2$ orbifold convention.}
\begin{equation}
S_h = \frac{\mathcal A}{2 G_5} = 4 \pi M_5^3 \left( \frac{r_h}{L} \right)^3 V_3 \,,
  \end{equation}
where $G_5 \equiv 1/(8\pi M_5^3)$ is the 5D Newton constant and $V_3 = \int d^3x$ the comoving volume.  The entropy density per unit of comoving volume is thus
\begin{equation}
s_h = \frac{S_h}{V_3} = \frac{1}{2G_5} \left( \frac{r_h}{L} \right)^3 \,. \label{eq:sh}
  \end{equation}

\section{Spacetime Fluctuations}
\label{se:fluc}

We analyze the fluctuations of the 5D metric in the $\Mnu^\pm$ spacetimes. The spectrum of these fluctuations contains important bits of information. In particular it tells us about the stability of the $\Mnu^\pm$ spacetimes, and whether or not gravity decouples at low energy in each of them. 

In this section it is convenient to work in conformal coordinates. Notice that the properties of the spacetimes on each side of the brane are not symmetric, see Eq.\,\eqref{eq:M_cases}. 
For  $\nu \neq 1$,  $z$ is in $\mathbb{R}_+$,  thus we have a bounded interval  $z\in(0,z_b]$ to the left of the brane, but a half-bounded interval   $z\in[z_b,\infty)$ to the right of the brane. 
   This fact has consequences for the spectrum, as is shown further below. We study each region systematically.

We first detail in subsection \ref{se:gauge_fixing}  our approach to spacetime fluctuations. Along the process, we review some technical points that are usually left unexplained in the warped extra dimension literature, and correct a statement about gauge fixing. Additional material is collected in Apps.\,\ref{app:gauge-fixing} and \ref{app:quadratic_action}. 

\subsection{Parametrization and Gauge-Fixing}

\label{se:gauge_fixing}

The metric fluctuations have in general the form $g_{MN}+\tilde h_{MN}$. Invariance of the action under diffeomorphisms implies the gauge symmetry $\tilde h_{MN}\to \tilde h_{MN} +\nabla_M \xi_N + \nabla_N \xi_M$ that must be used to remove five unphysical degrees of freedom. If one works with a boundary effective action in which the fifth dimension is integrated out, the 5D gauge symmetry becomes  a 4D gauge symmetry plus St\"uckelberg transformations. Details on the gauge fixing are given in App.\,\ref{app:gauge-fixing}, see also \cite{Hinterbichler:2011tt}. Our specific gauge choice is given below.

We parametrize the fluctuation in the form $g_{MN}+\tilde h_{MN}\equiv e^{-2A(z)}(\eta_{MN}+2 M_5^{-3/2} h_{MN})$ for convenience.  
We plug the metric fluctuations into the general action \eqref{eq:action}. 
The fluctua\-tion of the Ricci scalar is computed via an identity using conformal rescaling to flat space $g_{MN}=e^{-2A(z)}\eta_{MN}$, which gives \be ^{(5)}R=e^{2A(z)} \left({}^{(5)}R_{\rm flat}  + 8 ^{(5)}\square A  -12 (\partial_M A)^2  \right) \,, \label{eq:conf_change} 
\ee  and the middle term is conveniently integrated by parts. On the right-hand side, contractions are done using the 5D Minkowski metric $\eta_{MN}$.

The  fluctuations of $^{(5)}R_{\rm flat}$ at quadratic order in $h_{MN}$ are well-known, see e.g.~\cite{carroll2003spacetime}. 
Before gauge-fixing, the expression contains the combination $^{(5)}R_{\rm flat}\supset \frac{1}{2}\left((\partial_5 h_{\mu\nu})^2- (\partial_5 h)^2\right)$, which gives rise to the 4D Fierz-Pauli mass term when one decomposes the 5D fields over a basis 
of 4 modes and integrate out over $z$ (see e.g. \cite{Fichet:2021xfn}). This is how the spectrum of massive gravitons appears in the boundary effective action.  

 $^{(5)}R_{\rm flat} $ contains a kinetic mixing of the longitudinal part of $h_{\mu\nu}$ with $h^{55}$,   $\partial_{\mu}h^{\mu\nu} \partial_\nu h^{55}$. This is removed by a suitable field redefinition $h_{\mu\nu} \to h_{\mu\nu} -\frac{1}{2} \eta_{\mu\nu} h_{55}$. 
Finally, we fix the gauge such that $ h_{\mu5}=0$ and $ h^\mu_\mu \equiv h=0$, leaving the traceless part of $ h_{\mu\nu}$ and $h_{55}$ as physical degrees of freedom (see App.\,\ref{app:gauge-fixing}).

The physical, diagonalized fluctuations of the metric {and the dilaton vev} take the final form
\begin{align}
 ds^2&=  e^{-2 A(z)} \left[ e^{-2F(x,z)} \left(\eta_{\mu\nu}+ \frac{2} {M^{3/2}_5} \h_{\mu\nu}(x,z)\right)dx^\mu dx^\nu+(1 + 2 F(x,z))^2 dz^2\right]  \,,  \nonumber \\
\bar \Phi(x,z) &= \bar \phi(z) +\bar\varphi(x,z)  \,, \label{eq:ds2fluc}
\end{align}
where $h_{\mu\nu}$  is traceless and $F\equiv \frac{1}{2 }M_5^{-3/2}h_{55}$. This matches the gauge-fixed form usually taken in the warped extra dimension literature, see e.g.~\cite{Csaki:2000zn}.

The extra terms in \eqref{eq:conf_change} do not produce a mass term for the $h_{\mu\nu}$ field. We show explicitly in App.\,\ref{app:quadratic_action} that the quadratic fluctuation of the volume form does not produce a mass term for $h_{\mu\nu}$ either. Hence $h_{\mu\nu}$ is massless in the 5D action.

When using the gauge-fixed metric \eqref{eq:ds2fluc} in the equations of motions of the $F$ and $\bar\varphi$ fluctuations, a constraint equation relating $F$ and $\bar\varphi$ appears,
\be
\bar \phi^\prime(z) \bar\varphi(x,z) = \left(\partial_z-2A'(z) \right)F(x,z)\,.
\label{eq:gauge_radion}
\ee
Details are given in App.\,\ref{app:EoM_canonical_radion}. 
This is consistent with the choices of parametrization and gauge made in \cite{Csaki:2000zn,Megias:2015ory}, see also \cite{Maldacena:2002vr} for related considerations.
We emphasize that \eqref{eq:gauge_radion} is not a gauge choice by itself, a statement that is sometimes found in the warped extra dimension literature.

There is thus one physical scalar mode that is a combination of $F$ and $\bar\varphi$,  usually referred to as the \textit{radion}. For our purposes, it is enough to focus on the $F$ fluctuation. We refer to $F$ as the radion  for simplicity.

\subsection{Graviton}

We introduce the graviton propagator  $\left\langle h_{\mu\nu}(x^M)h_{\rho\sigma}(x'^N)\right\rangle\equiv G^h_{\mu\nu,\rho\sigma}\left(x^M,x'^N\right)$. 
The propagator contains a tensor structure that is determined by our (partially unitary and traceless) gauge fixing. For our purposes of determining the spectrum, it is enough to focus on the identity part of the tensor structure. We define the propagator of the identity part of  $G^h_{\mu\nu,\rho\sigma}$, 
\be
G^h_{\mu\nu,\rho\sigma}(x,x') = G_h(x,x') 
\mathbb{I}_{\mu\nu,\rho\sigma}
+ \ldots   \,,
\label{eq:Gh}
\ee
with 
$\mathbb{I}_{\mu\nu,\rho\sigma}=
 \frac{1}{2}\left(\eta_{\mu\rho}\eta_{\nu\sigma} + \eta_{\mu\sigma} \eta_{\nu\rho} \right) 
$ the identity on the space of 4D  symmetric tensors.
The wave operator for $h_{\mu\nu}$ is  read from the identity term in the fluctuation of the Einstein-Hilbert action, 
\be
\S_{{\rm EH}}= \frac{1}{2}\int d^5x \, e^{-3A} \left( h_{\mu\nu} {\cal O}^{\mu\nu,\rho\sigma} h_{\rho\sigma} \right) +\ldots  \,, \label{eq:SE_grav}
\ee
with 
\be
{\cal O}^{\mu\nu,\rho\sigma} = \mathbb{I}^{\mu\nu,\rho\sigma} {\cal D}\,,\quad\quad\quad {\cal D} = e^{3A}\partial_z\left(e^{-3A} \partial_z\right)+  \square^{(4)} \,.
\ee
The $G_h$ propagator in \eqref{eq:Gh} satisfies a  scalar equation of motion, 
\be
{\cal D} G_h(x,x') = i e^{3A}\delta^{(5)}(x-x')\,.
\label{eq:Gh_EOM}
\ee

Plugging the metric solutions and Fourier transforming along the Minkowski slices, the wave operator is 
\be
{\cal D} = \partial^2_z +\frac{3}{\nu^2-1} \frac{1}{z}\partial_z-p^2 \,,
\ee
with $p^2 = \eta_{\mu\nu} p^\mu  p^\nu$.
The homogeneous solutions of \eqref{eq:Gh_EOM} are any combination of $z^\alpha{I_\alpha}(\sqrt{p^2}z)$, $z^\alpha{K_\alpha}(\sqrt{p^2}z)$, where $I_\alpha$, $K_\alpha$ are the modified Bessel functions with order
\be
\alpha = \frac{1}{2} \frac{(4-\nu^2)}{(1-\nu^2)}\,.
\ee
The order satisfies $\alpha\geq 2$ for $0\leq\nu<1$ and $\alpha< 0$ for $1 < \nu < 2$. The case $\nu=1$ has been analyzed  in \cite{Fichet:2023xbu} and is not repeated here.

\subsubsection{The $z\in(0,z_b]$ region}

The boundary condition of the graviton  on the brane is Neumann, $\partial_zG^h(z,z')|_{z=z_b}=0$. 
We require regularity of the solutions in the $z\to0$  limit. 
In this region it is convenient to use the basis of solutions {$f_\pm=z^\alpha I_{\pm\alpha}(\sqrt{p^2}z)$}, where $I_a(x)$ is the modified Bessel function of the first kind.\,\footnote{A different basis is required in case $\alpha$ is integer.}

{\bf Case $0\leq\nu<1$}. 
This is the ${\M}^+_{\nu<1}$ spacetime. In that case $f_+$ is the regular solution at $z\to0$, since $f_+/f_- \xrightarrow[z\to 0]{}0\,.$
The bulk propagator is given by 
\be
G_h(z,z^\prime;p) = -\frac{i}{C}  f_+(z_<) \left(f_-(z_>)- \frac{f_-'(z_b)}{f_+'(z_b)} f_+(z_>)\right)  \,,
\ee
where $z_<={\rm min}(z,z')$,  $z_>={\rm max}(z,z')$. The  
 $C$ constant is fixed by the Wronskian of $f_\pm$ via $f'_+ f_-- f'_- f_+\equiv Ce^{3A}$
(see \cite{Fichet:2019owx}).

In particular, the brane-to-brane propagator reduces to 
\begin{equation}
G_h(z_b,z_b;p) = - i |1 - \nu^2|^3 \eta^3 z_b^3 \frac{I_\alpha(\sqrt{p^2} z_b)}{I_{\alpha-1}(\sqrt{p^2} z_b)}  \frac{1}{\sqrt{p^2}}\,. \label{eq:Gh_nuSRp}
\end{equation}
In the low-momentum limit, \be 
G_h(z_b,z_b;p)\xrightarrow[p\to 0]{} -i
\eta^3 z_b^4 \frac{(1 - \nu^2)^4 }{4-\nu^2}\,.
\ee

{\bf Case $1<\nu<2$}. 
This is the  ${\M}^-_{\nu>1}$ spacetime. In that case $f_-$ is the regular solution at $z\to0$, since $f_-/f_+ \xrightarrow[z\to 0]{}0\,.$
The bulk propagator is given by
{
\be
G_h(z,z^\prime;p) = -\frac{i}{C}  f_-(z_<) \left(f_+(z_>)- \frac{f_+'(z_b)}{f_-'(z_b)} f_-(z_>)\right)\,.
\ee
}
The brane-to-brane propagator is
\begin{equation}
G_h(z_b,z_b;p) = - i |1 - \nu^2|^3 \eta^3 z_b^3 \frac{I_{-\alpha}(\sqrt{p^2} z_b)}{I_{1-\alpha}(\sqrt{p^2} z_b)}  \frac{1}{\sqrt{p^2}}\,. \label{eq:Gh_nuLRm}
\end{equation}
In the low-momentum limit, \be 
G_h(z_b,z_b;p)\xrightarrow[p\to 0]{} -i \eta^3 z_b^2 \frac{(\nu^2-1)^2(\nu^2+2)}{p^2} \,.
\ee

{\bf Spectrum}. We can read the spectrum from the non-analyticities of  the propagators (\textit{e.g.}~Eqs.\,\eqref{eq:Gh_nuSRp} and \eqref{eq:Gh_nuLRm}) in the $-p^2\in \mathbb{R}_+$ region.  In the present case, it is a tower of poles corresponding to the zeros of the Bessel $J$ functions. 
In both cases, the spectrum is discrete with spacing $m_n\sim \frac{n \pi}{z_b}$ at large $n$. 
This can be understood as a consequence of the fact that the  interval in $z$ is bounded. The wave functions of the massive modes tend to be localized towards the brane. 

There is however an important difference between the $\MnuP$ and  $\MnuM$  cases. 
At the level of the spectrum, it turns out that $\MnuM$ has a graviton massless mode, while $\MnuP$ does not, as can be seen from the low-energy limits. We relate this feature to the fact that, in the $\MnuM$ case, $z=0$ corresponds to the curvature singularity, while in the $\MnuP$ case, $z=0$ corresponds instead to asymptotically flat infinity in the $r$ coordinates.    The massless mode is not localized towards the brane.
The singularity in the $\MnuM$ case  has, somehow,  a confining  effect that gives the zero mode a finite normalization.
In contrast, for $\MnuP$, even though the interval is bounded, there is no physical boundary at $z=0$, and accordingly  no normalizable massless mode exists in the spectrum. 
 A similar phenomenon appears in the LD spacetime, see \cite{Fichet:2023xbu}.

\subsubsection{The $z\in[z_b,\infty)$ region}

The boundary condition of the graviton  on the brane is Neumann, $\partial_zG^h(z,z')|_{z=z_b}=0$. 
We require regularity of the solutions in the $z\to\infty $  limit. 
In this region it is convenient to use the basis of solutions $\left\{ i_\alpha(z), k_\alpha(z) \right\} \equiv \left\{z^\alpha I_\alpha(\sqrt{p^2}z),z^\alpha K_{\alpha}(\sqrt{p^2}z)\right\}$, where $K_\alpha(x)$ is the modified Bessel function of the second kind.

In that region, for either $\nu$ smaller or larger than one, the regular solution at $z\to\infty$ is $k_\alpha$. The bulk propagator is thus 
\be 
G_h(z,z^\prime;p) = -\frac{i}{C} \left( i_\alpha(z_<) - \frac{i_\alpha'(z_b)}{k_\alpha'(z_b)} k_\alpha(z_<)  \right)k_\alpha(z_>) \,.
\ee
The brane-to-brane propagator is 
\begin{equation}
G_h(z_b,z_b;p) = -i |1 - \nu^2|^3 \eta^3 z_b^3 \frac{K_\alpha(\sqrt{p^2} z_b)}{K_{\alpha-1}(\sqrt{p^2} z_b)}  \frac{1}{\sqrt{p^2}}\,. \label{eq:Gh_nuLRp}
\end{equation}
We analyze the small momentum limit for both cases of $\nu$. 

 {\bf Case $0\leq\nu<1$}. 
This is the ${\M}^-_{\nu<1}$ spacetime. 
The brane-to-brane propagator {in the small momentum regime} is
\begin{equation}
G_h(z_b,z_b;p) = -i  \eta^3 z_b^4 \frac{|1 - \nu^2|^3}{2}
\left(
\frac{p^2 z_b^2}{4\alpha-4}+ \frac{\Gamma(1-\alpha)}{\Gamma(\alpha)}\left(\frac{\sqrt{p^2}z_b}{2}\right)^{2\alpha}
\right)^{-1}\left(1+O(p^2)\right)\,. \label{eq:Gh_nuMsmallp}
\end{equation}

{\bf Case $1<\nu<2$}. This is the  ${\M}^+_{\nu>1}$ spacetime. 
The brane-to-brane propagator {in the small momentum regime} is
{
\begin{equation}
G_h(z_b,z_b;p) = -i  \eta^3 z_b^4 \frac{|1 - \nu^2|^3}{2}
\left(
-\alpha +  \frac{\Gamma(\alpha+1)}{\Gamma(-\alpha)}\left(\frac{\sqrt{p^2}z_b}{2}\right)^{-2\alpha}
\right)^{-1}\left(1+O(p^2)\right)\,. \label{eq:Gh_nuPsmallp}
\end{equation}
}

{\bf Spectrum}. In both cases, the spectrum features an ungapped continuum, described  
by the leading non-analytical terms $p^{2\alpha}$ and  {$p^{-2\alpha}$}.
In the $\nu<1$ case, the spectrum also features a massless mode. This mode is not a ghost since $\alpha\geq 2$ on  $\nu\in [0,1)$. 
This mode is absent in the $\nu>1$ case. We thus obtain again that 
the existence of the zero mode is tied to the space being bounded by the curvature singularity.

{\bf Holography}. 
The case of AdS$_5$ spacetime truncated by a UV brane, which is the familiar background to study AdS/CFT, is recovered from the $\MnuM$ region when taking $\nu=0$. In that case, the form of the propagator \eqref{eq:Gh_nuMsmallp} is interpreted as a 4D free  field mixing with a CFT operator with dimension $\Delta=2+\alpha=4$, which is the dimension of the 4D CFT stress-energy tensor. 
One may notice that for $\nu\neq0$, the structure of \eqref{eq:Gh_nuMsmallp} remains unchanged. Therefore, if one speculatively generalizes the holographic interpretation to any $0\leq \nu<1$, we may interpret \eqref{eq:Gh_nuMsmallp} as a 4D graviton coupled to a 4D stress tensor with dimension 
\be
\Delta=2+\alpha  \,
\ee
in the dual theory. Since $\alpha>2$ for $0<\nu<1$, this dual stress tensor is an irrelevant operator.  
It would be very interesting to find a suitably deformed 4D CFT that reproduces such a behavior. 

{\bf Braneworld}. A $\MnuM$ braneworld scenario with $\nu<1$ could be viable  due to the existence of the graviton zero mode. For non-integer $\alpha$, the graviton continuum induces a modification to the Newtonian potential of the form 
{
\be \Delta V_{\rm Newton}(R) \sim \frac{1}{R}\Delta(R),\quad \Delta(R)=\left(\frac{z_b}{R}\right)^{2\alpha-2} \label{eq:VN_deviation} \ee 
}
(see e.g. \cite{Callin:2004py, Fichet:2022xol} for  calculations in AdS and LD cases). Notice that the AdS case ($\alpha=2$) is recovered from \eqref{eq:VN_deviation}, but recovering the LD case is not possible from this simple estimate because a mass gap  must emerge in the LD limit. 
Similar deviations with non-integer powers have been pointed out in the non-gravitational case  in AdS background \cite{Brax:2019koq,Costantino:2019ixl,Chaffey:2021tmj}. In the gravitational case, having a non-AdS background is mandatory to reach a non-integer scaling.   

{
Constraints from Newtonian gravity at distances $R\sim 25\, \mu m$~\cite{Smullin:2005iv} lead, after imposing the condition $\Delta(R)\lesssim 1$, to the mild bound
\begin{equation}
    \frac{1}{z_b}\gtrsim 0.01 \textrm{ eV}  \,.
    \label{eq:newtonianbound}
\end{equation}
}

\subsection{Radion }

We repeat our analysis for the radion fluctuation $F$.  Due to the gauge fixing \eqref{eq:gauge_radion}, we know that, even though $F$ is not the physical radion, it does have the same spectrum as the physical one.\,\footnote{We have verified explicitly this feature at the level of the radion effective action. 
The full radion effective action is derived in\,\cite{radion_paper}.}
We thus use $F$ as a proxy to find the radion spectrum. 
We refer to $F$ as the radion in the following. 

The wave operator for the radion is \cite{Csaki:2000zn,Megias:2015ory}
\be
{\cal D}F= e^{A}X \partial_z\left( \frac{e^{A}}{X} \partial_z( e^{-2A}  F ) \right)+ \left(\square^{(4)} - 2X\right)  F    \,  \label{eq:DF}
\ee
with \be X(z)=A''(z)+(A'(z))^2 \,.\ee
From this homogeneous EOM, we can  deduce  the equation for the radion propagator
$\langle F(x) F(x')\rangle  \equiv G_F(x,x')$ up to a positive normalization constant $c$,\,\footnote{
The dependence in $e^{3A} X$ is enforced  by the single-$\partial_z$ term of the EOM. This term determines the Wronskian of the solutions, which for consistency must be proportional to the $\delta^{(5)}(x-x')$ term \cite{Fichet:2019owx}. 
Here this is more easily derived by studying the EOM of the $\tilde F\equiv e^{-2A} F$ field. The Wronskian of the $\tilde F$ EOM is found to be $\propto e^{-A} X$. Going back from $G_{\tilde F}$ to $G_{ F}$ then produces the $e^{3A} X$ factor  used in Eq.\,\eqref{eq:GF_EOM}. }
\be
{\cal D} G_F (x,x')  = i c\, e^{3A} X \delta^{(5)}(x-x')\,. 
\label{eq:GF_EOM}
\ee
{Putting in the background and going to momentum space, we have}
\be
 {\cal D}= \partial_z^2  + \frac{(1+2\nu^2)}{(\nu^2-1)} \frac{1}{z} \partial_z  -p^2\,. \label{eq:GF_EOM_D}
\ee
The homogeneous solutions of \eqref{eq:GF_EOM} are any combination of $z^\gamma{I_\gamma}(\sqrt{p^2}z)$, $z^\gamma{K_\gamma}(\sqrt{p^2}z)$  with Bessel order 
\be
\gamma= \frac{1}{2} \frac{(2+\nu^2)}{(1-\nu^2)}=\alpha-1\,.
\ee
It is convenient to introduce
\begin{equation}
\Y_\pm \equiv 2 \left[ (1 - \nu^2)^2  \eta z_b^2 \frac{p^2}{U_b^{\prime\prime}} \pm 1 \right] \,,
\end{equation}
where $U_b^{\prime\prime} \equiv U_b^{\prime\prime}(v_b)$ is the second derivative of the brane-localized effective potential defined by Eq.~(\ref{eq:Ub_def}). The mass dimension is $[U_b^{\prime\prime}] = 1$.

\subsubsection{The $z\in(0,z_b]$ region}

The boundary condition of the radion on the brane is 
\begin{equation}
\left(\partial_z -   2A' + \frac{2 p^2 e^{A}}{U_b^{\prime\prime}(\phi)} \right) G_F(z,z') \Bigg|_{z=z_b} =0\,.  \label{eq:EoMF}
\end{equation}
We require regularity in the $z\to 0$ limit. A convenient basis of solutions is then  
$\left\{ i_\gamma(z) , \bar i_\gamma(z) \right\} \equiv \left\{z^\gamma I_\gamma(\sqrt{p^2}z),z^\gamma I_{-\gamma}(\sqrt{p^2}z)\right\}$.

{\bf Case $0\leq\nu<1$}.  This is the $\mathcal M_{\nu < 1}^+$ spacetime. {In that case the regular solution at $z \to 0$ is $i_\gamma$, since $i_\gamma/\bar i_{\gamma} \xrightarrow[z\to 0]{}0\,.$ The bulk propagator is given by
\be 
G_F(z,z^\prime;p) = -\frac{i }{C} i_\gamma(z_<) \left( \bar i_\gamma(z_>) - \frac{\bar i_\gamma^\prime(z_b) + \mathcal X_+ \bar i_\gamma(z_b)}{i_\gamma^\prime(z_b) + \mathcal X_+ i_\gamma(z_b)} i_\gamma(z_>)
\right) \,,
\ee
where $\mathcal X_+ \equiv -2 A^\prime(z_b) + \frac{2 p^2}{U_b^{\prime\prime}} e^{A(z_b)}$, and the constant $C$ is fixed by $ i_\gamma^\prime \bar i_\gamma  - \bar i_\gamma^\prime i_\gamma  \equiv C \, c \,  e^{3A} X$.}

The brane-to-brane propagator reduces to
\be
G_F(z_b,z_b;p) = -i \frac{Z }{ \Y_- + |1-\nu^2|  \frac{I_{\gamma-1}\left(  \sqrt{p^2} z_b \right)}{ I_\gamma\left( \sqrt{p^2} z_b \right) }\sqrt{p^2} z_b }  \, \label{eq:GF_nuSRp}
\ee
with $Z = c \nu^2 (1-\nu^2)^2 \eta^3 z_b^2$. Since our focus is on the spectrum, the precise value of $Z$ is irrelevant, we only need to know that $Z>0$. The same constant appears below.

In the low-momentum limit, an isolated mode appears with mass
\be
m_0^2 = \frac{\nu^2 (4-\nu^2)U_b^{\prime\prime}}{(1-\nu^2)^2[U_b^{\prime\prime}+ 2 (4 - \nu^2)\eta] z_b^2 }\,.
\ee
This mode becomes massless in the limits $\nu \to 0$ or $U''_b\to 0$.

{\bf Case $1<\nu<2$}. This is  the $\mathcal M_{\nu > 1}^-$ spacetime. {In that case the regular solution at $z \to 0$ is $\bar i_\gamma$. The bulk propagator is given by
\be 
G_F(z,z^\prime;p) = -\frac{i }{C} \bar i_\gamma(z_<) \left(  i_\gamma(z_>) - \frac{i_\gamma^\prime(z_b) + \mathcal X_+  i_\gamma(z_b)}{\bar i_\gamma^\prime(z_b) + \mathcal X_+ \bar i_\gamma(z_b)} \bar i_\gamma(z_>)
\right) \,.
\ee
}
{The $C$ constant is fixed by $ \bar i_\gamma^\prime  i_\gamma  -  i_\gamma^\prime \bar i_\gamma  \equiv C \, c \, e^{3A} X$. 
}
The brane-to-brane propagator is
{
\begin{equation}
G_F(z_b,z_b;p) = -i \frac{Z}{ \Y_+ + |1-\nu^2|  \frac{I_{1-\gamma}\left(  \sqrt{p^2} z_b \right)}{ I_{-\gamma}\left( \sqrt{p^2} z_b \right) }\sqrt{p^2} z_b }   \,. \label{eq:GF_nuLRm}
\end{equation}
}

In the low-momentum limit, an   isolated mode appears with mass
\be
m_0^2 = \frac{6\nu^2U_b^{\prime\prime}}{(\nu^2-1)^2[U_b^{\prime\prime} + 6\nu^2\eta]z_b^2}\,.
\ee

{\bf Spectrum}. In both cases, the spectrum features a discretum with spacing $m_n\sim \frac{n\pi}{z_b}$, as in the graviton case. 
The spectrum features, in either cases, a stable massive isolated mode which is non-ghost and non-tachyonic provided that $U_b^{\prime\prime} >0$.

\subsubsection{The $z\in[z_b,\infty)$ region}

The boundary condition of the radion on the brane is 
{
\begin{equation}
\left(\partial_z -  2A' - \frac{2 p^2 e^{A}}{U_b^{\prime\prime}(\phi)} \right) G_F(z,z') \Bigg|_{z=z_b} =0\,.  \label{eq:EoMF}
\end{equation}
}
We require regularity in the $z\to \infty$ limit. A convenient basis of solutions is then 
$\left\{ i_\gamma(z), k_\gamma(z) \right\} \equiv   \left\{z^\gamma I_\gamma(\sqrt{p^2}z),z^\gamma K_{\gamma}(\sqrt{p^2}z)\right\}$. {In that case, for either $\nu$ smaller of larger than one,  the regular solution at $z \to \infty$ is $k_\gamma$. The bulk propagator is given by
\be 
G_F(z,z^\prime;p) = -\frac{i }{C} \left( i_\gamma(z_<) - \frac{i_\gamma^\prime(z_b) + \mathcal X_- i_\gamma(z_b)}{k_\gamma^\prime(z_b) + \mathcal X_- k_\gamma(z_b)} k_\gamma(z_<)
\right) k_\gamma(z_>) \,,
\ee
where $\mathcal X_- \equiv -2 A^\prime(z_b) - \frac{2 p^2}{U_b^{\prime\prime}} e^{A(z_b)}$. } { The  
 $C$ constant is fixed by   $\pm \left( i'_\gamma k_\gamma - k'_\gamma i_\gamma \right) \equiv C \, c \, e^{3A} X$ } {for $\mathcal M^{\mp}$.}

{\bf Case $0\leq\nu<1$}. {This is the $\mathcal M_{\nu < 1}^-$ spacetime.} The brane-to-brane propagator is
{
\begin{equation}
G_F(z_b,z_b;p) =  - i \frac{Z }{ \Y_+ + |1-\nu^2|  \frac{K_{\gamma-1}\left(  \sqrt{p^2} z_b \right)}{ K_{\gamma}\left( \sqrt{p^2} z_b \right) }\sqrt{p^2} z_b} \,. \label{eq:GF_nuLRp}
\end{equation}
}

There is an isolated mode, that is non-ghost and non-tachyonic provided that $U_ b^{\prime\prime} \geq 0$. 
The mass is 
\be
m_0^2 = \frac{6\nu^2 U_b^{\prime\prime}}{(1-\nu^2)^2[U_b^{\prime\prime}  +6\nu^2\eta]z_b^2}\,.  \label{eq:m02_3}
\ee
When $\nu$ approaches $0$, appropriate Bessel approximation has to be used. We find that  the mass is nonzero at $\nu=0$,  $m_0^2 \simeq  \frac{U_b^{\prime\prime}}{\eta z_b^2}$.

{\bf Case $1<\nu<2$}. {This is the $\mathcal M_{\nu > 1}^+$ spacetime.} The brane-to-brane propagator is
{
\begin{equation}
G_F(z_b,z_b;p) =  - i \frac{ Z }{ \Y_- + |1-\nu^2|  \frac{K_{\gamma-1}\left(  \sqrt{p^2} z_b \right)}{ K_{\gamma}\left( \sqrt{p^2} z_b \right) }\sqrt{p^2} z_b} \,. \label{eq:GF_nuLRp}
\end{equation}
}
There is an isolated mode  
with mass
\be
m_0^2 = \frac{ \nu^2 (4-\nu^2)U_b^{\prime\prime}}{(\nu^2-1)^2 \left[ U_b^{\prime\prime} + 2 (4 - \nu^2)\eta \right] z_b^2}\,.
\ee

{\bf Spectrum}. In both cases, the spectrum features an ungapped continuum. The spectrum features also an isolated mode
that mixes with the continuum and acquires a decay width due to this mixing. The isolated mode is not stable, it can decay into the continuum. 
It  is non-ghost and non-tachyonic  if $U_b^{\prime\prime}>0$.

\subsubsection{Radion Mode and Stability}

We recapitulate the properties of the isolated radion mode. 
The radion mode exists in all regions and is non-ghost and non-tachyonic   if 
\be
U''_b>0\,. \label{eq:stab_U}
\ee
For $\nu\to 0$  the radion mode becomes massless in $\MnuP$ but remains massive in $\MnuM$. 
We also find that the radion wavefunction is localized towards the brane in $\MnuP$, while it has flat profile in $\MnuM$. Thus in $\MnuM$  the singularity acts somehow as a second boundary that allows the isolated mode to exist, analogously to the graviton case.

The study of the  isolated radion mode  is the  extension  of the holographic potential analysis of section \ref{se:stab}. While in section \ref{se:stab} we obtain only a stability condition, here we determine the actual mass of the radion fluctuation.
We find that the mass of the mode is described by  the same expression in the ${\cal M}_{\nu<1}^+$ and ${\cal M}_{\nu>1}^+$ cases, and similarly in the ${\cal M}_{\nu<1}^-$ and ${\cal M}_{\nu>1}^-$ cases.   Even though these masses are respectively proportional to $V_b^{\prime\prime} \mp  W^{\prime\prime}$, the fact that the expressions match exactly at $\nu<1$ and $\nu>1$ is unexpected, especially  because the various spacetime regions and  the rest of the  spectra are very different.   A full computation of the radion effective action might be needed to fully understand this unexpected simplicity.  We leave this for future work.

\section{Holographic Fluids}
\label{se:fluid}

We consider the $\Mnu$ spacetime with a planar black hole and a brane.  The black hole is always in the $\Mnu^-$ region, as shown in section \ref{se:model}.

\subsection{Effective Einstein Equation}

We study the gravitational behavior of the theory projected onto the brane.
The induced metric on the brane {at $r=r_b$} is
\begin{equation}
d\bar s^2 = \bar g_{\mu\nu} dx^\mu dx^\nu = -dt^2 + e^{-2A(r_b)} d\x^2 \,,  \label{eq:metric_brane}
\end{equation}
where the brane proper time is $dt = e^{-A(r_b)} \sqrt{f(r_b)} \, d\tau$. Notice the $e^{-A(r_b)}$ amounts to a spatial scale factor.

To study gravity in the holographic theory, we compute  the effective $4$-dimensional Einstein equations as seen by a brane observer. 
These are computed by  projecting the $5$-dimensional Einstein equations  on the brane using the Gauss equation and the Israel junction condition (see \cite{Shiromizu:1999wj} for the original calculation in AdS$_5$). We recall that  we use a $\mathbb{Z}_2$ orbifold convention as in \cite{Shiromizu:1999wj}
which implies that the spacetime is mirrored on the other side of the brane.  A notable implication is that the entropy of the black hole horizon is doubled when using this convention, see {e.g.}~Ref.~\cite{Megias:2018sxv}.

Following \cite{Fichet:2023xbu},  we find that the effective Einstein equations have the form
\begin{equation}
{}^{(d)}G_{\mu\nu} = \frac{1}{M_{4}^{2}} \left( T_{\mu\nu}^b + T_{\mu\nu}^{\rm eff}\right) + O\left( \frac{T_b^2}{M_5^{6}} \right) \,, \label{eq:D_1_Einstein}
\end{equation}
where $T_{\mu\nu}^b$ is the stress tensor of possible brane-localized  matter. 
The indices in \eqref{eq:D_1_Einstein} are contracted with the induced metric \eqref{eq:metric_brane}. Equation \eqref{eq:D_1_Einstein} has the form of the standard Einstein equations with an {extra} effective stress tensor $T^{\rm eff}_{\mu\nu}$. 
Moreover, it turns out that the structure of the $T^{\rm eff}_{\mu\nu}$ tensor represents a 
$4$-dimensional {perfect fluid} at rest:
\begin{equation}
T^{\rm eff,\mu}_\nu = g^{\mu\lambda} T^{\rm eff}_{\lambda \nu} =  \textrm{diag}(-\rho_{\rm eff}, P_{\rm eff}, \cdots, P_{\rm eff}) \,. 
\label{eq:Teff_gen}
\end{equation}
We refer to this as the holographic fluid. 
See \cite{Fichet:2023xbu} for further details. 

The effective energy density and pressure   in \eqref{eq:Teff_gen} split as
\begin{equation}
\rho_{\rm eff} = \rho_{\rm fluid} + \rho_{\rm vacuum} \,, \qquad P_{\rm eff} = P_{\rm fluid} + P_{\rm vacuum}  \,. \label{eq:rho_P_eff}
\end{equation}
The vacuum contributions  are independent on  $\nu$, $r_b$ and $r_h$. They are given by
\begin{equation}
\rho_{\rm vacuum} = -P_{\rm vacuum} = \Lambda_4 M_4^2 \,.
\end{equation}
The 4D cosmological constant $\Lambda_4$ and Planck mass $M_4$ are related to the parameters of the bulk action as
\begin{align}
\Lambda_4 &= -3 \eta^2 + \frac{\Lambda_b^2}{12 M_5^6} \,,  \label{eq:Lambda_4} \\
M_5^3 &= M_4^2  \eta \sqrt{1 + \frac{\Lambda_4}{3 \eta^2} } \,. \label{eq:M5M4}
\end{align}

The fluid contributions depend on $\nu$, $r_b$ and $r_h$. They are
\begin{align}
  \rho_{\rm fluid}(r_h,r_b) &= 3\eta^2 M_4^2 \left(  \frac{r_h}{r_b}\right)^{4-\nu^2} \,, \label{eq:rho_fluid} \\
  P_{\rm fluid}(r_h, r_b) &= (1-\nu^2) \eta^2 M_4^2 \left(  \frac{r_h}{r_b}\right)^{4-\nu^2} \,. \label{eq:P_fluid}
  \end{align}

For $\nu=0$, $\rho_{\rm fluid}$ and $P_{\rm fluid}$ scale as $r_b^{-4}$ and we have $\rho_{\rm fluid}=3 P_{\rm fluid}$, as expected from AdS$_5$. For $\nu=1$,  the pressure vanishes identically and 
$\rho_{\rm fluid}$ scales as $r_b^{-3}$, which is the scaling of pressureless matter. Once again, something peculiar happens for the LD$_5$ spacetime. 

Finally, for the critical case $\nu\to 2$, the $r_b$ dependence of
$\rho_{\rm fluid}$, $P_{\rm fluid}$  vanishes and we have $P_{\rm fluid}\to-\rho_{\rm fluid}$ asymptotically. 
Thus the fluid tends to behave as vacuum energy, i.e. as  a cosmological constant  for $\nu\to2$. In that limit, the separation  of the terms in \eqref{eq:rho_P_eff} does not hold. 
Accordingly, a direct calculation of the  $\nu=2$ case simply gives  $\rho_{\rm fluid}=0= P_{\rm fluid}$, because the black hole cannot exist  in this critical case. 

All these  results are summarized into  the equation of state of the holographic fluid, $P_{\rm fluid} = w_{\rm fluid} \,\rho_{\rm fluid}$ with parameter
 \begin{equation}
w_{\rm fluid} = \frac{1 - \nu^2}{3} \,. \label{eq:w_fluid}
 \end{equation}

\subsection{Thermodynamics}

We study the thermodynamics of the holographic fluid.
The physical parameters of the  system are the horizon location $r_h$ and the brane location $r_b$. We  treat all the thermodynamical variables as functions of these two parameters.

We define the volume and temperature of the system. These are related to the comoving volume $V_3=\int d^3x $ and to $r_h$ by powers of the scale factor
\begin{equation}
  a(r_b) \equiv e^{-A(r_b)} = \frac{r_b}{L} \,.
\end{equation}
 The spatial volume of the brane is given by
 \begin{equation}
V_b = V_3 \, a(r_b)^3 = V_3 \left( \frac{r_b}{L} \right)^3  \,.
   \end{equation}
  The black hole temperature on the brane is~\footnote{The temperature on the brane is obtained from the horizon temperature by multiplying with $1/\sqrt{|g_{\tau\tau}|}$. In the present analysis we are assuming $r_h \ll r_b$, hence we neglect the $1/\sqrt{f(r_b)}$ factor as it is a small correction.}
 \begin{equation}
T_b = \frac{T_h}{a(r_b)} = \frac{4 - \nu^2}{4} \frac{\eta}{\pi} \left( \frac{r_b}{r_h}\right )^{\nu^2 - 1} \,.  \label{eq:Tb}
   \end{equation}

Using the energy density obtained in \eqref{eq:rho_fluid}, 
 the total energy of the system  is
 \begin{equation}
E_{\rm fluid}(r_h,r_b) = \rho_{\rm fluid}(r_h,r_b) V_b = 3 \eta^2 M_4^2 \left( \frac{r_h}{L}\right)^3 \left( \frac{r_h}{r_b} \right)^{1-\nu^2} V_3 \,.
   \end{equation}
Using this expression of $E_{\rm eff}$ and using $T_b$ and $V_b$, we can derive the rest of the thermodynamic variables.

 By the fundamental laws of thermodynamics, the variation of the total energy of the system should satisfy
 \begin{equation}
dE = T dS - P dV \,.
   \end{equation}
 From this, we get the relation
 \begin{eqnarray}
\left[ \frac{\partial E_{\rm fluid}}{\partial r_h} -  T_b \frac{\partial  S_{\rm fluid}}{\partial r_h} \right] d r_h   + \left[ \frac{\partial E_{\rm fluid}}{\partial r_b} + P_{\rm fluid} \frac{\partial V_b}{\partial r_b}  - T_b \frac{\partial S_{\rm fluid}}{\partial r_b} \right]d r_b  = 0 \,. \label{eq:dE_fluid}
   \end{eqnarray}
Each bracket must separately vanish. From the first bracket, we find that
 \begin{equation}
S_{\rm fluid}(r_h,r_b) = 4\pi \eta M_4^2 \left( \frac{r_h}{L} \right)^3 V_3 + C_S(r_b) \,, \label{eq:S_eff}
   \end{equation}
 where $C_S(r_b)$ is an arbitrary function on $r_b$. We set $C_S$ to zero using that $S_{\rm eff}$ should vanish in the $r_h \to 0$ limit, for which the black hole does not exist. Plugging this result into the second bracket, we obtain the pressure, which matches exactly  
 the expression obtained in Eq.~(\ref{eq:P_fluid}). This provides  a nontrivial consistency check of our thermodynamic approach.

 Finally, the free energy of the system is
 \begin{equation}
F_{\rm fluid} \equiv E_{\rm fluid} - T_b S_{\rm fluid} = -(1-\nu^2) \eta^2 M_4^2 \left( \frac{r_h}{L} \right)^3 \left( \frac{r_h}{r_b} \right)^{1-\nu^2} V_3  \,,
\label{eq:free-energy}
   \end{equation}
which coincides with $-P_{\rm eff} V_b$.

These results make clear that the thermodynamics of the linear dilaton spacetime is very particular. For $\nu=1$, we have $F_{\rm fluid}=0$,  $T_b=$cste and $S_{\rm fluid}\propto E_{\rm fluid}$, which is the so-called Hagedorn behavior. See \cite{Fichet:2023xbu} for more details.

 We notice that the fluid entropy $S_{\rm fluid}$ can  be independently derived from the Bekenstein-Hawking entropy of the black hole.  Starting from the entropy density $s_h$ obtained  in \eqref{eq:sh}, we multiply by the redshift factor to get the entropy density in the brane, \textit{i.e.} $s_b = s_h/a(r_b)^3$. Multiplying with $V_b$, using the relation between the Planck masses of Eq.~(\ref{eq:M5M4}) and assuming $\Lambda_4 \ll \eta^2$, the result precisely reproduces Eq.~(\ref{eq:S_eff}):
 \begin{equation}
 s_b V_b = 4 \pi M_5^3 \left( \frac{r_h}{L} \right)^3 V_3 = S_{\rm fluid} \,. \label{eq:EOS}
   \end{equation}
In other words, we find $S_{\rm fluid} = S_h$, i.e. the entropy of the holographic fluid matches exactly the black hole entropy. 

In analogy with the matching of the entropies, we may expect that an appropriately defined mass density of the planar black hole, computed for example along the lines of \cite{Chang:2023kkq},  could also reproduce $\rho_{\rm fluid}$ upon redshifting. This is an interesting calculation that is left for future research.

 \subsection{Time Evolution of the Holographic Fluid}

We let the brane location evolve in time, $r_b=r_b(t)$. 
{Computing the time evolution of the system serves as a consistency check for  our formal results. }
  This is also a development that can be used for the study of realistic braneworld models beyond AdS$_5$, see discussion in next subsection. 
  The low-energy regime $\rho_b\ll \eta^2 M^2_4$ implies $H\ll \eta$ and the brane motion is nonrelativistic. We introduce the scale factor $a_b(t)\equiv a(r_b(t))$ and the Hubble parameter $H \equiv \dot a_b(t)/a_b(t)$.

{In this subsection we assume that there is no brane-localized matter, $T^b_{\mu\nu}=0$. The braneworld is empty, apart from the contributions from the holographic fluid and the vacuum energy both encoded into $T^{\rm eff}_{\mu\nu}$. }
 Starting from the effective Einstein equation \eqref{eq:D_1_Einstein}, 
 the Friedmann equations on the brane are
 \begin{align}
 3 M_{4}^{2} H^2  \approx -\rho_{\rm eff}  \,,   \quad \quad \quad 
 6 M_4^2 \frac{a_b^{\prime\prime}(t)}{a_b(t)} \approx - \left( \rho_{\rm eff} + 3 P_{\rm eff} \right)   \,. \label{eq:Friedmann_Eq1}
\end{align}

 The solution of the Friedmann equations including both the perfect fluid contribution and the cosmological constant term is
\begin{equation}
a_b(t) = a_h \left[  (4-\nu^2) \frac{\eta}{2\gamma} \sinh(\gamma(t-t_\ast)) \right]^{\frac{2}{4-\nu^2}}   \,, \quad (0 \le\nu < 2)  \label{eq:ab_t}
\end{equation} 
where 
\begin{equation}
\gamma \equiv \frac{1}{2} (4 - \nu^2) \sqrt{ \frac{\rho_{\rm vacuum}}{3 M_4^2} }  \,,
\end{equation}
and $a_h \equiv r_h / L$ while $t_\ast$ is an integration constant that can be freely chosen.  

The solution Eq.~(\ref{eq:ab_t})  reproduces a power-law behavior for a fluid-dominated universe of the form
\begin{equation}
a_b(t) \propto t^{\frac{2}{3(1+w_{\rm fluid})}} \, \qquad\quad (\rho_{\rm vacuum} \ll \rho_{\rm fluid}(r_b)) \,
\label{eq:at}
\end{equation}
with $w_{\rm fluid}$ given by Eq.~(\ref{eq:w_fluid}). 
Also, Eq.~(\ref{eq:ab_t}) leads to an exponential behavior for a universe dominated by the cosmological constant 
\begin{equation}
a_b(t) \propto \exp(Ht)   \qquad\quad (\rho_{\rm vacuum} \gg \rho_{\rm fluid}(r_b))  \,, \label{eq:ab_exponential}
\end{equation}
with $H = \sqrt{\Lambda_4 / 3}$.

The conservation equation of the  bulk stress tensor evaluated on the brane can be written as \cite{Tanaka:2003eg,Langlois:2003zb} 
\begin{equation}
\dot \rho_{\rm eff} + 4 H \rho_{\rm eff} + H T_\mu^{{\rm eff}\, \mu} =  - 2 \left( 1 + \frac{\rho_b}{\Lambda_b}\right) T^\phi_{MN} u^M n^N \,. \label{eq:conservation}
\end{equation}
It turns out that $T^\phi_{MN} u^M n^N  = O(H^3)$, the rhs of this equation is negligible in the low-energy regime.  Using that $T_\mu^{{\rm eff}\, \mu}  = -\nu^2 \rho_{\rm fluid}(r_b) - 4 \rho_{\rm vacuum}$, it is easy to verify that the conservation equation is satisfied.

Finally, the limiting case $\nu = 2$ has vacuum contribution only, such that the solution of the Friedmann equations is given by Eq.~(\ref{eq:ab_exponential}). In this case $\rho_{\rm fluid}(r_b) = 0$, and the conservation equation is also satisfied.

\subsection{A Cosmological Dark Fluid}

{We may identify the brane-localized matter $T^b_{\mu\nu}$ in Eq.\,\eqref{eq:D_1_Einstein} with the known matter content of our universe. The $\MnuM$ model then defines a braneworld scenario. In the low-energy regime, the main cosmological consequence is the presence of the holographic fluid. }

{
In the LD$_5$ case the fluid may be identified as dark matter \cite{Fichet:2022ixi,Fichet:2022xol}. 
But more generally for  $\nu\neq 1$, given the equation of state established in \eqref{eq:w_fluid}, 
 the behavior of the ``dark fluid'' can lie anywhere  between radiation-like  and vacuum energy-like. Radiation corresponds to $\nu=0$, i.e. $w_{\rm fluid}=\frac{1}{3}$, which is the well-known AdS$_5$ case. Vacuum energy behavior corresponds to the $\nu\to 2$ limit, i.e. $w_{\rm fluid}=-1$. 
This range of possibilities implies that the dark fluid may possibly dominate at some intermediate phase of the evolution of the universe. 
Here we do not undertake the study of this  intriguing set of possibilities --- a detailed analysis of each  cases would deserve a separate work. }

{
As a simple constraint, we just require that the fluid energy density be negligible at present times. This constraint, together with the one from Newtonian gravity shown in (\ref{eq:newtonianbound}), translates as  bounds on the fundamental parameters of the $\MnuM$ braneworld, that we now present.    }

{
The relationship between conformal and cosmological coordinates leads to
\begin{equation}
    |1-\nu^2|\eta z_b=\frac{L}{r_b(t)} \,,
\end{equation}
and makes $z_b=z_b(t)$. The bound (\ref{eq:newtonianbound}) on $z_b$ refers to the present time $t_0$, for which we define $a_b(t_0)=r_b(t_0)/L=1$. This translates into the {mild} bound on the $\eta$ parameter
\begin{equation}
    \eta=\frac{1}{|1-\nu^2|z_b(t_0)}\gtrsim \frac{0.01 \textrm{ eV}}{|1-\nu^2|} \,,
\end{equation}
for $\nu\neq 1$. The case $\nu=1$ has been treated in \cite{Fichet:2022xol}. 

{
Moreover, the fraction of fluid energy density in the universe $\Omega_{\rm fluid}=\rho_{\rm fluid}(t)/\rho_c(t)$, where the critical energy density is defined as $\rho_c=3 H^2 M_4^2$, yields
\begin{equation}
    \Omega_{\rm fluid}=\frac{\eta^2}{H^2}\, \left(\frac{r_h}{r_b}\right)^{4-\nu^2} \,. 
\end{equation}
We  then impose the condition that the fluid energy density be negligible with respect to the other known cosmological fluid  densities today:  $\Omega_{\rm fluid}<\Omega_{\rm rad}\approx 10^{-4}$. 
This in turn provides the bound
\begin{equation}
    \frac{r_h}{L}\lesssim \left(0.01\frac{H_0}{\eta}\right)^{\frac{2}{4-\nu^2}} \lesssim 10^{-\frac{66}{4-\nu^2}}
\end{equation}
where $H_0$ is the Hubble parameter today, which leads \textit{e.g.}~to the bound  {$r_h\lesssim 5 \times 10^{-17}L$ } for $\nu=0$, and $r_h\lesssim 10^{-22}L$ for $\nu=1$. 
}

\section{Summary}
\label{se:summary}

We present a simple class of 5D dilaton-gravity spacetimes that includes both AdS$_5$ and LD$_5$ (the linear dilaton spacetime) as particular cases.
The model has a single continuous parameter $\nu$. 
We compute the planar black hole solutions of these  $\Mnu$ spacetimes.

Depending on the $\nu$ parameter, the $\Mnu$ spacetimes can have a timelike boundary that is either conformal or regular.  $\Mnu$ also typically features a curvature singularity, except for certain values of $\nu$.  The  $\nu$ parameter is bounded (namely $\nu\in [0,2)$) for the black hole to be allowed to exist. The black hole horizon screens the singularity.

We assume that a flat brane exists,  that 
splits the $\Mnu$ spacetime  into inequivalent regions $\Mnu^\pm$.
The $\MnuM$ region contains the singularity and the black hole. 
The dilaton vev is stabilized on the brane by a potential $V_b(\phi)$. 
We compute the holographic potential of the dilaton-brane system and determine a stability condition on $V_b$ that is conveniently 
expressed using a brane-localized effective potential $U_b$. 
We show that the brane location can be free upon an appropriate tuning of the brane tension. The brane would otherwise go into the singularity or run to infinity.  This tuning turns out to be equivalent to setting the effective 4D cosmological constant to zero in the 4D holographic theory.

We perform a systematical analysis of the fluctuations of the bulk metric. 
Depending on the value of the $\nu$ parameter, the spectrum of the fluctuations for both gravitons and radion can be either discrete or continuous. Overall, the tendency of the spectral distribution, except for a possible isolated mode (see below), is to be localized in the direction of the curvature singularity for both $\MnuM$  and  $\MnuP$ spaces.

 The spectrum contains an isolated massless graviton mode in $\MnuM$. This implies that  low-energy 4D EFTs arising from $\MnuM$ feature  Einstein-like gravity.
 The  spectrum contains  an isolated massive radion mode in both  $\MnuM$  and  $\MnuP$, that is  non-tachyonic in all regions provided that the second derivative of the brane-localized effective potential  $U_b(\phi)$ is positive. This bound is equivalent to the one derived from the holographic potential stability, confirming  stability of the solutions upon  perturbations.
 Both graviton and radion modes  have flat profiles in $\MnuM$.  The singularity acts effectively as a second boundary that makes these modes  normalizable.

The elementary properties of the $\Mnu^\pm$ spacetimes are summarized in Table \ref{tab:summary}.
The class of $\Mnu$ spacetimes  sheds some light on the LD spacetime.
It is clear  from Table\,\ref{tab:summary}  that the LD spacetime  corresponds to a critical case between two  regimes, this helps to make sense of its  peculiar properties (see \cite{Fichet:2023xbu}). 
 \begin{table}[t]
    \centering
    \resizebox{15cm}{!}{
    \begin{tabular}{|c|c|c|c|c|c|c|c|}
    \hline
       Value of $\nu$  & $0$ [AdS$_5$] & $(0,1)$ & $1$ [LD$_5$] & $(1,\nu_c)$ & $\nu_c$ & $(\nu_c,2)$ & $2$  \\
       \hline\hline
   Scalar curvature   & \multicolumn{4}{c|}{$<0$}    & $0$   & \multicolumn{2}{c|}{$>0$}   \\      
                \hline
    Singularity  & \xmark  & \multicolumn{3}{c|}{good}  & \xmark & good & bad \\
   \hline
   Timelike boundary   & \multicolumn{2}{c|}{conformal}  & \xmark   & \multicolumn{4}{c|}{regular}    \\    
                \hline
    Bulk Black hole    & \multicolumn{6}{c|}{\cmark}  & \xmark \\
                \hline
    Gravity spectrum in ${\cal M}_\nu^-$    & \multicolumn{2}{c|}{continuum}  & gapped cont. & \multicolumn{4}{c|}{discretum} \\
                         \hline                       
    Gravity spectrum in ${\cal M}_\nu^+$     & \multicolumn{2}{c|}{discretum} & gapped cont. & \multicolumn{4}{c|}{continuum} \\         
         \hline
    \end{tabular}}
    \caption{Elementary properties of the $\Mnu$ spacetimes. The $\nu = \nu_c$ case {with $\nu_c \equiv \sqrt{5/2}$} is not Ricci-flat. There is always a massless graviton  mode in ${\cal M}_\nu^-$, and never in  ${\cal M}_\nu^+$. There is an isolated massive radion mode anywhere in ${\cal M}_\nu^\pm$, which is resonant when the spectrum is continuous. 
    }
    \label{tab:summary}
\end{table}

We compute the effective Einstein equation on the brane, which at low-energies features a  perfect fluid that is the manifestation of the bulk   black hole  projected on the brane.  
We find that the equation of state of the  holographic fluid is $w = \frac{1-\nu^2}{3}$. 
This interpolates between radiation behavior (AdS), non-relativistic behavior (LD), and tends to vacuum energy behavior for $\nu\to 2$. We verify the consistency of these results using the 5D conservation equation. { As another check we compute the time evolution of the fluid-dominated braneworld. }

{
The $\MnuM$ background provides a set of braneworld models beyond AdS$_5$. Its  main cosmological prediction is the existence of a dark fluid whose equation of state can lie anywhere between radiation ($\nu=0$) and vacuum energy ($\nu\to 2$) behaviors. The dark fluid could dominate at some intermediate period in the history of the universe. This generates a set of intriguing modifications of the standard cosmological history
that require dedicated investigation. Here we just derived a few simple bounds on the parameters of the $\MnuM$ braneworld. 
}

We establish the thermodynamic properties of the holographic fluid. It turns out that the entropy matches exactly the black hole Bekenstein-Hawking entropy upon appropriate redshifting. Along the same line of black hole/fluid correspondence,  it would be good to evaluate the black hole mass density using the formalism  of \cite{Chang:2023kkq} to verify whether it reproduces $\rho_{\rm fluid}$ upon redshifting.

{At the technical level, we explain in details the gauge-fixing of spacetime fluctuations in warped spacetimes, and present 
detailed calculations of the graviton action. We provide a detailed computation of the graviton quadratic action, using a form of the off-shell gravity action that is reminiscent of the Codazzi equation.  We show that the graviton mass induced by the variation of the $\sqrt{g}$ volume form vanishes. We also revisit the equations of motion of the radion/dilaton sector, showing that a redundancy in the equations of motion fixes an integration constant, which in turn completely fixes the relation between the metric and dilaton fluctuations, leaving the radion as the only degree of freedom. 
}

Regarding future research directions, it would be {for example} interesting to study a more evolved field content of the $\Mnu$ spacetimes. In particular, given the simplicity of  the $\Mnu$ spacetimes solutions,   it would be interesting to  exactly solve  the Abelian Higgs model introduced in \cite{Chaffey:2023xmz} in the $\Mnu$  background, in which case the Higgs field is identified as the dilaton field. 
Along the same lines, it would also be interesting to study models that solve the electroweak hierarchy problem using the $\nu\ll1$ limit, for which the Higgs/dilaton spectrum features a parametrically light mode. 

Finally it would be good to further study the holographic correlators defined on the brane. 
In this  paper we have only evaluated the two-point brane-to-brane propagators. We note that in the region $\MnuM$ with $\nu<1$, which is connected to the familiar case of AdS$_5$ with a UV brane, the two-point correlator  takes a  form which amounts to  a simple generalization of the well-known AdS result.
It would be very interesting to  further explore the  brane correlators of $\MnuM$ and to search for a dual 4D theory that reproduces their features.

\begin{acknowledgments}

We would like to thank Hooman Davoudiasl and Tony Gherghetta for discussions. The work of EM is supported by the project PID2020-114767GB-I00 and by the Ram\'on y Cajal Program under Grant RYC-2016-20678 funded by MCIN/AEI/10.13039/501100011033 and by ``FSE Investing in your future'', by Junta de Andaluc\'{\i}a under Grant FQM-225, and  by the ``Pr\'orrogas de Contratos Ram\'on y Cajal'' Program of the University of Granada. The work of MQ is partly supported by Spanish MICIN under Grant PID2020-115845GB-I00, and by the Catalan Government under Grant 2021SGR00649. IFAE is partially funded by the CERCA program of the Generalitat de Catalunya.
\end{acknowledgments}

\appendix

\section{Gauge-Fixing Spacetime Fluctuations  }

\label{app:gauge-fixing}

In curved space, the 5D fluctuation $h_{MN}$ of the $g_{MN}$ metric  has the gauge redundancy
\be
h_{MN}\to h_{MN} + \nabla_M \xi_N + \nabla_N \xi_M \,,
\label{eq:gauge_gen}
\ee
due to invariance under diffeomorphisms. 

In the case of warped metrics of the general form \eqref{eq:ds2brane}, one can always define a decomposition over a set of orthogonal wavefunctions $h_{MN}(x,z)=\sum_\lambda h^\lambda_{MN}(x) f^{\lambda}(z)$. In the presence of a brane, a brane-localized mode can also exist, see \cite{Fichet:2021xfn} for details on the decomposition.  For our purposes, it is enough to focus on a single mode $f^\lambda$. The gauge parameter is similarly expanded as $\xi_{M}(x,z)=\sum_\lambda \xi^\lambda_{M}(x) f^{\lambda}(z)$. The gauge transformation \eqref{eq:gauge_gen} can be written as 
\begin{eqnarray}
h^\lambda_{\mu\nu} &\to& h^\lambda_{\mu\nu} + \nabla_\mu \xi^\lambda_\nu + \nabla_\nu \xi^\lambda_\mu\,, \label{eq:gauge_munu}  \\
h^\lambda_{5\mu} &\to& h^\lambda_{5\mu} + \omega^\lambda \xi^\lambda_\mu + \nabla_\mu \xi^\lambda_5\,, \label{eq:gauge_5mu}  \\
h^\lambda_{55} &\to& h^\lambda_{55} + 2\omega^\lambda \xi^\lambda_5
\label{eq:gauge_55}  \,,
\end{eqnarray}
where $\omega^\lambda = (f^\lambda)^{-1}\nabla_5 f^\lambda $. 
The $\omega^\lambda$ may be a function of $z$, however the gauge transformation is done within the action, in which the $z$ integral is ultimately done. 
For the present discussion we can just use \eqref{eq:gauge_munu}-\eqref{eq:gauge_55} assuming that $\omega^\lambda$ is constant. 
Hence the situation is analogous to gauge fixing spacetime fluctuations in e.g.~a  $S_1$ compactification \cite{Hinterbichler:2011tt}: the 5D gauge symmetry becomes a 4D Stueckelberg symmetry.

\paragraph{The full unitary gauge.}
The gauge redundancy \eqref{eq:gauge_munu}-\eqref{eq:gauge_55} can be used to set $h^\lambda_{5\mu}=0$ by fixing $\xi_\mu^\lambda$, and to set $h^\lambda_{55}=0$ by fixing $\xi_5$. This is the full unitary gauge, in which all polarizations of the massive graviton modes are encoded into the propagator of $h^\lambda_{\mu\nu}$.  
Notice that the trace of $h^\lambda_{\mu\nu}$ cannot be set to zero.

\paragraph{The traceless-unitary gauge.}
Instead of removing $h^\lambda_{55}$ we can remove the  4D trace $h^\lambda=h^\lambda_{\mu\nu}g^{\mu\nu}$, along with  $h^\lambda_{5\mu}$, as follows. 
The 4D trace transforms as $h^\lambda\to h^\lambda +2\nabla^\mu\xi^\lambda_\mu$. We see that $\nabla^\mu\xi^\lambda_\mu$ is related to the longitudinal component of $\xi^\lambda_\mu$, {that we will call $\xi^\lambda_{\mu,L}$}, {which can then be used to remove $h^\lambda$}. Let us then introduce the transverse and longitudinal projectors 
\be
P_{T,\nu}^\mu = \delta^{\mu}_\nu - \frac{\nabla^\mu \nabla_\nu}{\nabla^2} \,,\quad\quad P_{L,\nu}^\mu =  \frac{\nabla^\mu \nabla_\nu}{\nabla^2} \,,  \label{eq:P_TL}
\ee
that satisfy $P_{T\,\alpha}^\mu P^\alpha_{L,\nu} =0$, $P^\mu_{T, \alpha} P^{\alpha}_{T,\nu} = P^\mu_{T,\nu}$, and similarly for $P_L$.  We apply $P_T$ to \eqref{eq:gauge_5mu} and use  that,  
in the general warped metric, $\nabla_\mu$ and $\nabla_5$ commute since $R_{\mu 5}=0$. The $\xi^\lambda_5$ term is projected out, leaving
\be  h^\lambda_{5\mu,T}\to  h^\lambda_{5\mu,T} + \omega^\lambda \xi^\lambda_{\mu,T} \,, 
\ee 
with $h^\lambda_{5\mu,T}\equiv P_T\cdot h^\lambda_{5\mu}$, $\xi^\lambda_{\mu,T}\equiv P_T\cdot \xi^\lambda_{\mu}$. 
Using this transformation we remove the 
transverse component of $h^\lambda_{5\mu}$ using the transverse component of $\xi_\mu$.  
We \textit{cannot} remove the longitudinal part of $h^\lambda_{\mu5}$ because the corresponding longitudinal part of $\xi^\lambda_\mu$ has been already used to remove the trace. 
However, we can use $\xi^\lambda_5$, since the longitudinal part transforms as
\be 
 h^\lambda_{5\mu,L}\to h^\lambda_{5\mu,L} + \omega^\lambda   \xi^\lambda_{\mu,L}  +\nabla_\mu \xi^\lambda_5 \,.  
\ee
In doing so, $h_{5\mu}$ is completely removed.  All gauge freedoms have been used, thus $h_{55}$ remains unconstrained. 

In summary we have traded $h_{55}^\lambda=0$ for $h^\lambda=0$. In this traceless and partially unitary gauge, the $h^\lambda_{\mu\nu}$ propagator contains spin-$2$ and spin-$1$ polarizations, while the spin-$0$ polarization 
of the massive graviton mode is separately encoded into the propagator of~$h^\lambda_{55}$.

\section{Quadratic Action and Graviton Mass }
\label{app:quadratic_action}

By considering the expansion of the action up to second order in the metric fluctuations, we  prove in this appendix the absence of mass terms for the graviton. This is a point that is often left implicit in the warped extra dimension literature, where fluctuations of the $\sqrt{g}$ volume form  at second order are often ignored.

\subsection{Codazzi Form of the Action}
\label{subapp:off_shell}

Let us consider the 5D bulk metric in proper coordinates
\begin{equation}
ds^2 = \gamma_{\mu\nu} dx^\mu dx^\nu + dy^2 \,,  \qquad \gamma_{\mu\nu} = e^{-2A(y)} \eta_{\mu\nu} \,.
\end{equation}
The non-vanishing Christoffel symbols are~\cite{Landsteiner:2011iq}
\begin{equation}
-\Gamma^{5}_{\mu\nu} = K_{\mu\nu} = \frac{1}{2} \dot \gamma_{\mu\nu} \,, \qquad \Gamma^{\mu}_{\nu 5} = K^\mu{}_\nu \,,
\end{equation}
where the dot denotes differentiation with respect to $y$. We consider intrinsic four dimensional curvature quantities on the $y = $ constant surface. The 5D Ricci scalar writes in terms of 4D quantities as
\begin{equation}
{}^{(5)}R = {}^{(4)}R - 2 \dot K - K^2 - K_{\mu\nu} K^{\mu\nu} \,, \label{eq:R_5D}
\end{equation}
with ${}^{(4)} R^{\mu}{}_{\nu\alpha\beta}$ the 4D Riemann tensor. In the 4D quantities, indices are raised and lowered with $\gamma_{\mu\nu}$, e.g. $K = \gamma^{\mu\nu} K_{\mu\nu}$, ${}^{(4)} R = \gamma^{\mu\nu} \, {}^{(4)} R_{\mu\nu}$. After plugging Eq.~(\ref{eq:R_5D}) into the action of Eq.~(\ref{eq:action}), we integrate by parts the term $\propto \dot K$, thus producing a boundary term that exactly cancels the GHY term. The final form of this \textit{off-shell} action writes
\begin{eqnarray}
\S &=& \int d^5x \sqrt{g} \left( \frac{M_5^3}{2} \left(  {}^{(4)} R  +  K^2 - K_{\mu\nu} K^{\mu\nu}  \right)  - \frac{1}{2} (\partial_\mu \phi)^2 - \frac{1}{2} \dot \phi^2  - V(\phi)  \right)  \nonumber \\
&&-\int_{\textrm{brane}} d^4x \sqrt{\gamma} \left(V_b(\phi) + \Lambda_b \right) ~ + ~ \S_{\textrm{matter}} \,.
\end{eqnarray}
We remind that in this manuscript we have adopted the orbifold convention in which space is mirrored on each side of the brane, as in e.g.~\cite{Shiromizu:1999wj}.

\subsection{Action at Second Order}
\label{subapp:action_2nd_order}

Let us allow for fluctuations in the 4D part of the metric,
\begin{equation}
\gamma_{\mu\nu} \equiv e^{-2A(y)} \bar \gamma_{\mu\nu} \,, \qquad  \bar \gamma_{\mu\nu}  = \eta_{\mu\nu} + h_{\mu\nu}   \,.
\end{equation}
We are fixing the gauge such that $h_{\mu 5} = 0$ and $h^\mu{}_\mu \equiv h = 0$. We will omit in this analysis the dilaton vev, i.e. $h_{55} = 0 = \bar\varphi$. Both metrics $\gamma_{\mu\nu}$ and $\bar\gamma_{\mu\nu}$ are related by a Weyl rescaling with scaling function $\omega \equiv A(y)$ which is constant in $x^\mu$.  The Ricci scalar ${}^{(4)} R$ is thus  expressed in terms of ${}^{(4)} {\bar R}$ (the Ricci scalar computed with the metric $\bar \gamma_{\mu\nu}$) as
\begin{equation}
{}^{(4)}  R  = e^{2A(y)} \, {}^{(4)}  {\bar R}  \,.
\end{equation}
On the other hand, the extrinsic curvature computed up to second order ${\mathcal O}(h_{\mu\nu}^2)$ leads to the following results:
\begin{eqnarray}
K &=& -4 \dot A - \frac{1}{2} h_{\mu\nu } \partial_5 h^{\mu\nu} \,, \qquad \\
K_{\mu\nu} K^{\mu\nu} &=& 4 \dot A^2 + \dot A h_{\mu\nu} \partial_5 h^{\mu\nu} + \frac{1}{4} \partial_5 h_{\mu\nu}  \partial_5 h^{\mu\nu} \,. \label{eq:K}
\end{eqnarray}
Finally, after using the classical equations of motion for the background, the on-shell action becomes
\begin{eqnarray}
\S  &=&   \int d^5 x \, \sqrt{\gamma} \left[ \frac{M_5^3}{2}  \left( e^{2A(y)}  \,  {}^{(4)}  {\bar R}  + 3 A^\prime(y) h_{\mu\nu} \partial_5 h^{\mu\nu} - \frac{1}{4} \partial_5 h_{\mu\nu} \partial_5 h^{\mu\nu} \right)  - 2 V(\phi) \right]  \nonumber \\
&& - \int_{\textrm{brane}} d^4x \sqrt{\gamma} \Lambda_b + \S_{\textrm{matter}}   \,,  \label{eq:S5}
\end{eqnarray}
where we have set the minimum value of the brane potential to zero by convention, i.e. $V_b(v_b) = 0$.

\subsubsection{Brane terms}

Let us study the brane term. The contribution of this term to the energy-momentum tensor is
\begin{equation}
  T_{\mu\nu} = - \Lambda_b \gamma_{\mu\nu} \,.  \label{eq:Smunu}
\end{equation}
By using Israel's junction condition, we have~\cite{Shiromizu:1999wj}
\begin{equation}
[K_{\mu\nu}] = - \frac{1}{M_5^3} \left( T_{\mu\nu} - \frac{1}{3} \gamma_{\mu\nu} T \right) \,, 
\end{equation}
where $T \equiv \gamma^{\mu\nu} T_{\mu\nu}$ and $[X] \equiv X^+ - X^-$, while the $\mathbb Z_2$ symmetry implies that $K^+_{\mu\nu} = - K^-_{\mu\nu}$. Focussing on $K^-_{\mu\nu}$ we have 
\begin{equation}
K^- \equiv \gamma^{\mu\nu} K_{\mu\nu}^- = \frac{2}{3 M_5^3}  \Lambda_b \,,  \label{eq:K2}
\end{equation}
and from a comparison with Eq.~(\ref{eq:K}) we obtain
\begin{equation}
\Lambda_b = - 6 M_5^3 A^\prime(y_b) \,.  \label{eq:Lambda_b}
\end{equation}
This result can be used to evaluate the brane tension contribution to the action, i.e.
\begin{eqnarray}
  \S_b &\equiv& - \int_{\textrm{brane}} d^4x  \sqrt{\gamma} \Lambda_b =  6 M_5^3 \int_{\textrm{brane}} d^4x \sqrt{\gamma} A^\prime(y_b)  \,.  \label{eq:Sb}
\end{eqnarray} 
The volume form has the following expansion in fluctuations
\begin{equation}
\sqrt{\gamma} = e^{-4A(y)} \left( 1 - \frac{1}{4} h_{\mu\nu} h^{\mu\nu}  + \cdots \right) \,.
\end{equation}
We  use this to work out the expression of $\S_b$ at orders ${\cal O}(h_{\mu\nu}^0)$ and ${\cal O}(h_{\mu\nu}^2)$. 

\paragraph{Order ${\cal O}(h_{\mu\nu}^0)$.}

It is convenient to rewrite the boundary terms via integration by parts, i.e. via the divergence theorem. 
At ${\cal O}(h_{\mu\nu}^0)$ we have
\begin{eqnarray}
\sqrt{\gamma} A^\prime(y) |_{y_b} &=& - \frac{1}{4}   \frac{d}{dy} \left( e^{-4A(y)}\right) \Big|_{y_b} + {\cal O}(h_{\mu\nu}^2) = - \frac{1}{4}  \int^{y_b} dy  \frac{d^2}{dy^2} \left( e^{-4A(y)}\right) + {\cal O}(h_{\mu\nu}^2) \nonumber \\
&=& \frac{2}{3M_5^3} \int^{y_b} dy  \, e^{-4A(y)} \,V(\phi)  + {\cal O}(h_{\mu\nu}^2)   \,, \label{eq:gA}
\end{eqnarray}
where   we have used the background equations of motion in the last equality. 
We also used that $\partial_y\left( e^{-4A(y)}\right)$ is vanishing on other possible boundaries --- for example at the singularity in the $\Mnu$ background \eqref{eq:ds2_Mnu}.  
The brane contribution at lowest order in the fluctuation takes therefore the form
\begin{equation}
\S_b = \int d^5x \, e^{-4A(y)} 2 V(\phi) + {\cal O}(h_{\mu\nu}^2)  \,. \label{eq:SV}
\end{equation}
This  exactly cancels the bulk term $\propto V(\phi)$ in Eq.~(\ref{eq:S5}) at this order.~\footnote{The integral in Eq.~(\ref{eq:SV}) has been extended to the full domain considering the $\mathbb Z_2$ symmetry and the subsequent factor $1/2$.}

\paragraph{Order ${\cal O}(h_{\mu\nu}^2)$.}

In the computation at order ${\cal O}(h_{\mu\nu}^2)$, we have to consider $\S_b$ as well as the bulk terms $\propto \sqrt{\gamma} V(\phi)$ and $\propto  \sqrt{\gamma} A^\prime(y) h_{\mu\nu} \partial_5 h^{\mu\nu}$  in the action (\ref{eq:S5}). The terms are
\begin{eqnarray}
\S_b^{(2)} &\equiv&    6 M_5^3 \int_{\textrm{brane}} d^4x [\sqrt{\gamma}]_{2} A^\prime(y_b)   = -\frac{3}{2} M_5^3 \int_{\textrm{brane}} d^4x \, e^{-4 A(y_b) }A^\prime(y_b) h_{\mu\nu} h^{\mu\nu} \,, \\
\S_V^{(2)} &\equiv&  \int d^5 x \, [\sqrt{\gamma}]_{2}   \left( - 2V(\phi) \right) = \frac{1}{2} \int d^5 x \, e^{-4A(y)} V(\phi) h_{\mu\nu} h^{\mu\nu}   \,, \\
\S_{A^\prime}^{(2)} &\equiv&    \frac{3}{2} M_5^3 \int d^5 x \, [\sqrt{\gamma}]_{0}  A^\prime(y) h_{\mu\nu} \partial_5 h^{\mu\nu}  =  \frac{3}{4} M_5^3 \int d^5 x \, e^{-4A(y)}  A^\prime(y) \partial_5 \left( h_{\mu\nu}  h^{\mu\nu} \right) \,, \nonumber \\
\label{eq:SAp}
\end{eqnarray}
respectively, where $[\sqrt{\gamma}]_{n}$ stands for the order ${\cal O}(h_{\mu\nu}^n)$ term of the volume form. After integrating by parts in Eq.~(\ref{eq:SAp}) with the $\mathbb Z_2$ orbifold  convention, it turns out that the summation of all these contributions cancel out, i.e. $\S_b^{(2)} + \S_V^{(2)} + \S_{A^\prime}^{(2)} = 0$.

We conclude from this computation that there is no mass term for the fluctuation~$h_{\mu\nu}$. The final result for the on-shell quadratic action of $h_{\mu\nu}$ is 
\begin{equation}
\S =  \frac{M_5^3}{2} \int d^5x  \, e^{-2A(y)} 
 \left[  -\frac{1}{4} \partial_\rho h_{\mu\nu} \partial^\rho h^{\mu\nu} + \frac{1}{2} \partial_\rho h_{\mu\nu} \partial^\mu h^{\rho\nu}   - \frac{1}{4} e^{-2A(y)} 
 \partial_5 h_{\mu\nu} \partial_5 h^{\mu\nu} \right]  + \S_{\textrm{matter}}   \,.  \label{eq:SS5}
\end{equation}
The two first terms come from the expansion of ${}^{(4)} {\bar R}$.  
The action (\ref{eq:SS5}) matches the result presented in the warped extra dimension literature~\cite{Cabrer:2009we,Gherghetta:2010cj}.

\section{Equations of Motion of the Scalar Sector}
\label{app:EoM_canonical_radion}

The classical equations of motion for the $F$ and $\bar \varphi$ fluctuations are~\cite{Csaki:2000zn}
\begin{eqnarray}
 &&\square^{(4)} F(x,z) + \partial_z^2 F(x,z) - A^\prime(z) \partial_z F(x,z) - 2 \bar\phi^\prime(z) \partial_z \bar\varphi(x,z) = 0 \,,   \label{eq:EoM1}  \\
 &&\bar \phi^\prime(z) \partial_\mu\bar\varphi(x,z) - \left(\partial_z-2A^\prime(z) \right)\partial_\mu F(x,z) = 0  \,, \label{eq:EoM2der}  \\
&&\square^{(4)} \varphi(x,z) - \partial_z^2 \varphi(x,z) + 3 A^\prime(z) \partial_z \varphi(x,z) + e^{-2A} V^{\prime\prime}(\phi) \varphi(x,z) \nonumber \\
  &&\qquad + 6 \phi^\prime(z) \partial_z F(x,z) + 4 e^{-2A }V^\prime(\phi) F(x,z) = 0 \,.  \label{eq:EoM3}  
\end{eqnarray}
Eq.~(\ref{eq:EoM2der}) can be integrated  to give
\be
\bar \phi^\prime(z) \bar\varphi(x,z) - \left(\partial_z-2A^\prime(z) \right)F(x,z) = c(z)  \,, \label{eq:EoM2}
\ee
with $c(z)$ an arbitrary function constant in $x^\mu$.  
This constant is set to zero due to a combination of the two other equations \eqref{eq:EoM1}, \eqref{eq:EoM3}, which satisfy 
\begin{equation}
\left( 1 + c_1 \partial_z \right) [\textcolor{blue}{\ref{eq:EoM1}}] + \left( c_2 + c_3 \partial_z + c_4 \partial_z^2 \right) [\textcolor{blue}{\ref{eq:EoM2}}] + c_5 \,[\textcolor{blue}{\ref{eq:EoM3}}] = 0 \,,
\end{equation}
with
\begin{eqnarray}
c_1 &=&  \frac{1}{2} \left( \nu^2  - 1 \right) z \,, \qquad c_2 =  \frac{1}{2M_5^3} \left[ \frac{1+\nu^2}{1-\nu^2} + \frac{1}{3} (\nu^2-1 ) p^2 z^2 \right] \frac{1}{z}  \,, \\
c_3 &=&  \frac{1}{6 M_5^3} \,, \qquad c_4 =  \frac{1}{6 M_5^3} (\nu^2-1)  z  \,, \qquad c_5 = \frac{\nu}{2\sqrt{3 M_5^3}}  \,,
\end{eqnarray}
once $c(z)=0$.

As a result,  Eq.~(\ref{eq:EoM2}) amounts to a constraint equation which relates the fluctuations  $\bar\varphi(x,z)$ and  $F(x,z)$.
We are left with two independent differential equations, e.g. (\ref{eq:EoM1}) and (\ref{eq:EoM2}) with $c(z)=0$. 
The fact that redundancies in the equations of motions turn into nontrivial algebraic constraints on integration constants was observed in \cite{Fichet:2022ixi,Fichet:2023xbu} at the level of the background.

\bibliographystyle{JHEP}
\bibliography{biblio}

\providecommand{\href}[2]{#2}\begingroup\raggedright\begin{thebibliography}{10}

\bibitem{Aharony:1999ti}
O.~Aharony, S.~S. Gubser, J.~M. Maldacena, H.~Ooguri, and Y.~Oz, {\it {Large N
  field theories, string theory and gravity}},  {\em Phys. Rept.} {\bf 323}
  (2000) 183--386, [\href{http://arxiv.org/abs/hep-th/9905111}{{\tt
  hep-th/9905111}}].

\bibitem{Seiberg:1997zk}
N.~Seiberg, {\it {New theories in six-dimensions and matrix description of M
  theory on T**5 and T**5 / Z(2)}},  {\em Phys. Lett. B} {\bf 408} (1997)
  98--104, [\href{http://arxiv.org/abs/hep-th/9705221}{{\tt hep-th/9705221}}].

\bibitem{Berkooz:1997cq}
M.~Berkooz, M.~Rozali, and N.~Seiberg, {\it {Matrix description of M theory on
  T**4 and T**5}},  {\em Phys. Lett. B} {\bf 408} (1997) 105--110,
  [\href{http://arxiv.org/abs/hep-th/9704089}{{\tt hep-th/9704089}}].

\bibitem{Aharony:1998ub}
O.~Aharony, M.~Berkooz, D.~Kutasov, and N.~Seiberg, {\it {Linear dilatons, NS
  five-branes and holography}},  {\em JHEP} {\bf 10} (1998) 004,
  [\href{http://arxiv.org/abs/hep-th/9808149}{{\tt hep-th/9808149}}].

\bibitem{Aharony:1999ks}
O.~Aharony, {\it {A Brief review of 'little string theories'}},  {\em Class.
  Quant. Grav.} {\bf 17} (2000) 929--938,
  [\href{http://arxiv.org/abs/hep-th/9911147}{{\tt hep-th/9911147}}].

\bibitem{Kutasov:2001uf}
D.~Kutasov, {\it {Introduction to little string theory}},  {\em ICTP Lect.
  Notes Ser.} {\bf 7} (2002) 165--209.

\bibitem{Antoniadis:2021ilm}
I.~Antoniadis, C.~Markou, and F.~Rondeau, {\it {Aspects of compactification on
  a linear dilaton background}},  {\em JHEP} {\bf 09} (2021) 137,
  [\href{http://arxiv.org/abs/2106.15184}{{\tt arXiv:2106.15184}}].

\bibitem{Gubser:1999vj}
S.~S. Gubser, {\it {AdS / CFT and gravity}},  {\em Phys. Rev.} {\bf D63} (2001)
  084017, [\href{http://arxiv.org/abs/hep-th/9912001}{{\tt hep-th/9912001}}].

\bibitem{Shiromizu:1999wj}
T.~Shiromizu, K.-i. Maeda, and M.~Sasaki, {\it {The Einstein equation on the
  3-brane world}},  {\em Phys. Rev. D} {\bf 62} (2000) 024012,
  [\href{http://arxiv.org/abs/gr-qc/9910076}{{\tt gr-qc/9910076}}].

\bibitem{Binetruy:1999hy}
P.~Binetruy, C.~Deffayet, U.~Ellwanger, and D.~Langlois, {\it {Brane
  cosmological evolution in a bulk with cosmological constant}},  {\em Phys.
  Lett. B} {\bf 477} (2000) 285--291,
  [\href{http://arxiv.org/abs/hep-th/9910219}{{\tt hep-th/9910219}}].

\bibitem{Hebecker:2001nv}
A.~Hebecker and J.~March-Russell, {\it {Randall-Sundrum II cosmology, AdS /
  CFT, and the bulk black hole}},  {\em Nucl. Phys. B} {\bf 608} (2001)
  375--393, [\href{http://arxiv.org/abs/hep-ph/0103214}{{\tt hep-ph/0103214}}].

\bibitem{Langlois:2002ke}
D.~Langlois, L.~Sorbo, and M.~Rodriguez-Martinez, {\it {Cosmology of a brane
  radiating gravitons into the extra dimension}},  {\em Phys. Rev. Lett.} {\bf
  89} (2002) 171301, [\href{http://arxiv.org/abs/hep-th/0206146}{{\tt
  hep-th/0206146}}].

\bibitem{Langlois:2003zb}
D.~Langlois and L.~Sorbo, {\it {Bulk gravitons from a cosmological brane}},
  {\em Phys. Rev. D} {\bf 68} (2003) 084006,
  [\href{http://arxiv.org/abs/hep-th/0306281}{{\tt hep-th/0306281}}].

\bibitem{Fichet:2023xbu}
S.~Fichet, E.~Meg\'\i{}as, and M.~Quir\'os, {\it {Holography of Linear Dilaton
  Spacetimes from the Bottom Up}},  {\em Phys. Rev. D} {\bf 109} (2024), no.~10
  106011, [\href{http://arxiv.org/abs/2309.02489}{{\tt arXiv:2309.02489}}].

\bibitem{Giveon:2017nie}
A.~Giveon, N.~Itzhaki, and D.~Kutasov, {\it {$ \mathrm{T}\overline{\mathrm{T}}
  $ and LST}},  {\em JHEP} {\bf 07} (2017) 122,
  [\href{http://arxiv.org/abs/1701.05576}{{\tt arXiv:1701.05576}}].

\bibitem{Giribet:2017imm}
G.~Giribet, {\it {$T\bar{T}$-deformations, AdS/CFT and correlation functions}},
   {\em JHEP} {\bf 02} (2018) 114, [\href{http://arxiv.org/abs/1711.02716}{{\tt
  arXiv:1711.02716}}].

\bibitem{Asrat:2017tzd}
M.~Asrat, A.~Giveon, N.~Itzhaki, and D.~Kutasov, {\it {Holography Beyond AdS}},
   {\em Nucl. Phys. B} {\bf 932} (2018) 241--253,
  [\href{http://arxiv.org/abs/1711.02690}{{\tt arXiv:1711.02690}}].

\bibitem{Araujo:2018rho}
T.~Araujo, E.~O. Colg\'ain, Y.~Sakatani, M.~M. Sheikh-Jabbari, and
  H.~Yavartanoo, {\it {Holographic integration of $T \bar{T}$
  \textbackslash{}\& $J \bar{T}$ via $O(d,d)$}},  {\em JHEP} {\bf 03} (2019)
  168, [\href{http://arxiv.org/abs/1811.03050}{{\tt arXiv:1811.03050}}].

\bibitem{Chakraborty:2020fpt}
S.~Chakraborty, G.~Katoch, and S.~R. Roy, {\it {Holographic complexity of LST
  and single trace $ T\overline{T} $}},  {\em JHEP} {\bf 03} (2021) 275,
  [\href{http://arxiv.org/abs/2012.11644}{{\tt arXiv:2012.11644}}].

\bibitem{Chakraborty:2020yka}
S.~Chakraborty, {\it {$
  \frac{\mathrm{SL}\left(2,\mathrm{\mathbb{R}}\right)\times
  \mathrm{U}(1)}{\mathrm{U}(1)} $ CFT, NS5+F1 system and single trace $
  T\overline{T} $}},  {\em JHEP} {\bf 03} (2021) 113,
  [\href{http://arxiv.org/abs/2012.03995}{{\tt arXiv:2012.03995}}].

\bibitem{Georgescu:2022iyx}
S.~Georgescu and M.~Guica, {\it {Infinite $\mathrm{T\bar T}$-like symmetries of
  compactified LST}},  \href{http://arxiv.org/abs/2212.09768}{{\tt
  arXiv:2212.09768}}.

\bibitem{Chang:2023kkq}
C.-K. Chang, C.~Ferko, and S.~Sethi, {\it {Holography and irrelevant
  operators}},  {\em Phys. Rev. D} {\bf 107} (2023), no.~12 126021,
  [\href{http://arxiv.org/abs/2302.03041}{{\tt arXiv:2302.03041}}].

\bibitem{Chakraborty:2023mzc}
S.~Chakraborty, A.~Giveon, and D.~Kutasov, {\it {Comments on single-trace $
  T\overline{T} $ holography}},  {\em JHEP} {\bf 06} (2023) 018,
  [\href{http://arxiv.org/abs/2303.12422}{{\tt arXiv:2303.12422}}].

\bibitem{Chakraborty:2023zdd}
S.~Chakraborty, A.~Giveon, and D.~Kutasov, {\it {Momentum in Single-trace
  $T\bar T$ Holography}},  \href{http://arxiv.org/abs/2304.09212}{{\tt
  arXiv:2304.09212}}.

\bibitem{Aharony:2023dod}
O.~Aharony and N.~Barel, {\it {Correlation functions in $
  \textrm{T}\overline{\textrm{T}} $-deformed Conformal Field Theories}},  {\em
  JHEP} {\bf 08} (2023) 035, [\href{http://arxiv.org/abs/2304.14091}{{\tt
  arXiv:2304.14091}}].

\bibitem{Gubser:2000nd}
S.~S. Gubser, {\it {Curvature singularities: The Good, the bad, and the
  naked}},  {\em Adv. Theor. Math. Phys.} {\bf 4} (2000) 679--745,
  [\href{http://arxiv.org/abs/hep-th/0002160}{{\tt hep-th/0002160}}].

\bibitem{Cabrer:2009we}
J.~A. Cabrer, G.~von Gersdorff, and M.~Quir\'os, {\it {Soft-Wall
  Stabilization}},  {\em New J. Phys.} {\bf 12} (2010) 075012,
  [\href{http://arxiv.org/abs/0907.5361}{{\tt arXiv:0907.5361}}].

\bibitem{Mann:2009id}
R.~B. Mann and R.~McNees, {\it {Boundary Terms Unbound! Holographic
  Renormalization of Asymptotically Linear Dilaton Gravity}},  {\em Class.
  Quant. Grav.} {\bf 27} (2010) 065015,
  [\href{http://arxiv.org/abs/0905.3848}{{\tt arXiv:0905.3848}}].

\bibitem{Hinterbichler:2011tt}
K.~Hinterbichler, {\it {Theoretical Aspects of Massive Gravity}},  {\em Rev.
  Mod. Phys.} {\bf 84} (2012) 671--710,
  [\href{http://arxiv.org/abs/1105.3735}{{\tt arXiv:1105.3735}}].

\bibitem{carroll2003spacetime}
S.~Carroll, {\em Spacetime and Geometry: An Introduction to General
  Relativity}.
\newblock Benjamin Cummings, 2003.

\bibitem{Fichet:2021xfn}
S.~Fichet, {\it {On holography in general background and the boundary effective
  action from AdS to dS}},  {\em JHEP} {\bf 07} (2022) 113,
  [\href{http://arxiv.org/abs/2112.00746}{{\tt arXiv:2112.00746}}].

\bibitem{Csaki:2000zn}
C.~Csaki, M.~L. Graesser, and G.~D. Kribs, {\it {Radion dynamics and
  electroweak physics}},  {\em Phys. Rev. D} {\bf 63} (2001) 065002,
  [\href{http://arxiv.org/abs/hep-th/0008151}{{\tt hep-th/0008151}}].

\bibitem{Megias:2015ory}
E.~Meg\'\i{}as, O.~Pujol\`as, and M.~Quir\'os, {\it {On dilatons and the LHC
  diphoton excess}},  {\em JHEP} {\bf 05} (2016) 137,
  [\href{http://arxiv.org/abs/1512.06106}{{\tt arXiv:1512.06106}}].

\bibitem{Maldacena:2002vr}
J.~M. Maldacena, {\it {Non-Gaussian features of primordial fluctuations in
  single field inflationary models}},  {\em JHEP} {\bf 05} (2003) 013,
  [\href{http://arxiv.org/abs/astro-ph/0210603}{{\tt astro-ph/0210603}}].

\bibitem{Fichet:2019owx}
S.~Fichet, {\it {Braneworld effective field theories --- holography,
  consistency and conformal effects}},  {\em JHEP} {\bf 04} (2020) 016,
  [\href{http://arxiv.org/abs/1912.12316}{{\tt arXiv:1912.12316}}].

\bibitem{Callin:2004py}
P.~Callin and F.~Ravndal, {\it {Higher order corrections to the Newtonian
  potential in the Randall-Sundrum model}},  {\em Phys. Rev. D} {\bf 70} (2004)
  104009, [\href{http://arxiv.org/abs/hep-ph/0403302}{{\tt hep-ph/0403302}}].

\bibitem{Fichet:2022xol}
S.~Fichet, E.~Meg\'\i{}as, and M.~Quir\'os, {\it {Cosmological dark matter from
  a bulk black hole}},  {\em Phys. Rev. D} {\bf 107} (2023), no.~11 115014,
  [\href{http://arxiv.org/abs/2212.13268}{{\tt arXiv:2212.13268}}].

\bibitem{Brax:2019koq}
P.~Brax, S.~Fichet, and P.~Tanedo, {\it {The Warped Dark Sector}},  {\em Phys.
  Lett. B} {\bf 798} (2019) 135012,
  [\href{http://arxiv.org/abs/1906.02199}{{\tt arXiv:1906.02199}}].

\bibitem{Costantino:2019ixl}
A.~Costantino, S.~Fichet, and P.~Tanedo, {\it {Exotic Spin-Dependent Forces
  from a Hidden Sector}},  {\em JHEP} {\bf 03} (2020) 148,
  [\href{http://arxiv.org/abs/1910.02972}{{\tt arXiv:1910.02972}}].

\bibitem{Chaffey:2021tmj}
I.~Chaffey, S.~Fichet, and P.~Tanedo, {\it {Continuum-Mediated Self-Interacting
  Dark Matter}},  {\em JHEP} {\bf 06} (2021) 008,
  [\href{http://arxiv.org/abs/2102.05674}{{\tt arXiv:2102.05674}}].

\bibitem{Smullin:2005iv}
S.~J. Smullin, A.~A. Geraci, D.~M. Weld, J.~Chiaverini, S.~P. Holmes, and
  A.~Kapitulnik, {\it {New constraints on Yukawa-type deviations from Newtonian
  gravity at 20 microns}},  {\em Phys. Rev.} {\bf D72} (2005) 122001,
  [\href{http://arxiv.org/abs/hep-ph/0508204}{{\tt hep-ph/0508204}}]. [Erratum:
  Phys. Rev.D72,129901(2005)].

\bibitem{radion_paper}
S.~Barbosa, S.~Fichet, E.~Megias, and M.~Quiros, {\it {Entanglement and Thermal
  Transitions from Singularities}},
  \href{http://arxiv.org/abs/2406.02899}{{\tt arXiv:2406.02899}}.

\bibitem{Megias:2018sxv}
E.~Meg\'\i{}as, G.~Nardini, and M.~Quir\'os, {\it {Cosmological Phase
  Transitions in Warped Space: Gravitational Waves and Collider Signatures}},
  {\em JHEP} {\bf 09} (2018) 095, [\href{http://arxiv.org/abs/1806.04877}{{\tt
  arXiv:1806.04877}}].

\bibitem{Tanaka:2003eg}
T.~Tanaka and Y.~Himemoto, {\it {Generation of dark radiation in bulk inflaton
  model}},  {\em Phys. Rev. D} {\bf 67} (2003) 104007,
  [\href{http://arxiv.org/abs/gr-qc/0301010}{{\tt gr-qc/0301010}}].

\bibitem{Fichet:2022ixi}
S.~Fichet, E.~Meg\'\i{}as, and M.~Quir\'os, {\it {Continuum effective field
  theories, gravity, and holography}},  {\em Phys. Rev. D} {\bf 107} (2023),
  no.~9 096016, [\href{http://arxiv.org/abs/2208.12273}{{\tt
  arXiv:2208.12273}}].

\bibitem{Chaffey:2023xmz}
I.~Chaffey, S.~Fichet, and P.~Tanedo, {\it {Holography of broken U(1)
  symmetry}},  {\em JHEP} {\bf 05} (2024) 330,
  [\href{http://arxiv.org/abs/2309.00040}{{\tt arXiv:2309.00040}}].

\bibitem{Landsteiner:2011iq}
K.~Landsteiner, E.~Meg\'\i{}as, L.~Melgar, and F.~Pena-Benitez, {\it
  {Holographic Gravitational Anomaly and Chiral Vortical Effect}},  {\em JHEP}
  {\bf 09} (2011) 121, [\href{http://arxiv.org/abs/1107.0368}{{\tt
  arXiv:1107.0368}}].

\bibitem{Gherghetta:2010cj}
T.~Gherghetta, {\it {A Holographic View of Beyond the Standard Model Physics}},
   in {\em {Theoretical Advanced Study Institute in Elementary Particle
  Physics}: {Physics of the Large and the Small}}, pp.~165--232, 2011.
\newblock \href{http://arxiv.org/abs/1008.2570}{{\tt arXiv:1008.2570}}.

\end{thebibliography}\endgroup

\end{document}